\documentclass[12pt,reqno]{amsart}

\usepackage{setspace}
\usepackage{ulem}
\usepackage[T1]{fontenc}
\usepackage{ae,aecompl}
\usepackage{graphics,cancel,graphicx}
\usepackage{epsfig,psfrag,manfnt}
\usepackage{mathrsfs,amsmath}
\usepackage{graphicx,mathabx}
\usepackage{bbm,subfigure,rotating,bm,float,mathdots,wasysym,scalerel,color,xcolor}
\usepackage{mathrsfs,setspace}
\usepackage{graphics,cancel,graphicx,mathabx}
\usepackage{epsfig,psfrag,manfnt}
\usepackage{bbm,subfigure,rotating,bm,float,mathdots,wasysym,scalerel}
\usepackage{array}
\usepackage{comment}

\definecolor{cerulean}{rgb}{0.0, 0.48, 0.65}
\definecolor{burntorange}{rgb}{0.8, 0.33, 0.0}
\definecolor{green}{rgb}{0.3,0.6,0.3}
\definecolor{red}{rgb}{1, 0.2, 0.1}

\usepackage[T1]{fontenc}
\usepackage{ae,aecompl}
\usepackage{mathrsfs,setspace}
\usepackage{graphics,cancel,graphicx,mathabx}
\usepackage{epsfig,psfrag,manfnt}
\usepackage{bbm,subfigure,rotating,bm,float,mathdots,wasysym,scalerel,ulem,yfonts}
\usepackage{array}

\usepackage{relsize}
\usepackage{exscale,xparse}

\usepackage{tikz-cd}
\usepackage{tikz}
\usetikzlibrary{positioning} 

\usepackage{array}
\usepackage[hidelinks]{hyperref}
\usepackage{xcolor}
\hypersetup{
    colorlinks,
    linkcolor={burntorange!50!black},
    citecolor={blue!50!black},
    urlcolor={blue!80!black}
}
\makeatletter
\newcommand*\bigcdot{\mathpalette\bigcdot@{.5}}
\newcommand*\bigcdot@[2]{\mathbin{\vcenter{\hbox{\scalebox{#2}{$\m@th#1\bullet$}}}}}
\makeatother

\usepackage{tikz}
\usepackage[framemethod=TikZ]{mdframed}
\mdfsetup{skipabove=\topskip,skipbelow=\topskip}
\tikzstyle{titregris} =
    [draw=gray, thick, fill=gray,%
        text=black, rectangle,  
        right,minimum height=.7cm]
\pgfdeclarehorizontalshading{exercisebackground}{100bp}{color(0bp)=(gray!40);color(100bp)=(black!5)}
\makeatletter
\def\mdf@@exercisepoints{}
\define@key{mdf}{exercisepoints}{
    \def\mdf@exercisepoints{#1}
}
\def\mdf@@myboxedtitle{}
\define@key{mdf}{myboxedtitle}{
    \def\mdf@myboxedtitle{#1}
}
\mdfdefinestyle{exercisestyle}{%
    outerlinewidth=1em,%
    outerlinecolor=white,%
    leftmargin=-1em,%
    rightmargin=-1em,%
    middlelinewidth=1.2pt,%
    roundcorner=5pt,%
    linecolor=gray,%
    apptotikzsetting={\tikzset{mdfbackground/.append style={gray!30}}},
    innertopmargin=1.2\baselineskip,
    skipabove={\dimexpr0.5\baselineskip+\topskip\relax},
    skipbelow={-1em},
    needspace=3\baselineskip,
    singleextra={%
        \node[titregris,xshift=1cm] at (P-|O) %
            {~\mdf@frametitlefont{\mdf@myboxedtitle}~};
},%
    firstextra={%
\node[titregris,xshift=1cm] at (P-|O) %
{~\mdf@frametitlefont{\mdf@myboxedtitle}~};
},
}
\makeatother

\NewDocumentCommand{\qfrac}{smm}{%
  \dfrac{\IfBooleanT{#1}{\vphantom{\big|}}#2}{\mathstrut #3}%
}

\usepackage{adjustbox}

\usepackage{relsize}
\usepackage{exscale,xparse}

\usepackage{tikz-cd}
\usepackage{tikz}
\usetikzlibrary{positioning} 

\usepackage{array}

\usepackage{hyperref}

\usepackage[all,cmtip]{xy}
\usepackage{amssymb,amsmath,amsfonts,amsthm,color}

\newcommand{\calD}{\mathcal{D}}

\newcommand{\calL}{\mathcal{L}}

\newcommand{\calS}{\mathcal{S}}

\newcommand{\mC}{\mathbb{C}}

\newcommand{\mR}{\mathbb{R}}

\newcommand{\mZ}{\mathbb{Z}}

\newcommand{\bbg}{\mathbf{g}}
\newcommand{\bbh}{\mathbf{h}}

\newcommand{\bbE}{\mathbf{E}}

\newcommand{\bbT}{\mathbf{T}}
\newcommand{\bbU}{\mathbf{U}}

\newtheorem{theorem}{Theorem}[section]

\newtheorem{proposition}[theorem]{Proposition}

\theoremstyle{definition}

\newtheorem{remark}[theorem]{Remark}

\theoremstyle{definition}

\theoremstyle{definition}

\theoremstyle{definition}

\begin{document}



\vspace{-0.2cm}

 \title[]{Elastic rods and elastic spinning rings as gravitational wave detectors}
 
 \author[]{Jos\'e Nat\'ario}
 \address{CAMGSD, Departamento de Matem\'{a}tica, Instituto Superior T\'{e}cnico, Universidade de Lisboa, Portugal}
 \email{jnatar@math.ist.utl.pt}
 \author[]{Amol Sasane}
 \address{Mathematics Department, London School of Economics, Houghton Street, London WC2A 2AE, UK}
 \email{A.J.Sasane@LSE.ac.uk}
 \author[]{Rodrigo Vicente}
 \address{Institut de Fisica d'Altes Energies (IFAE), The Barcelona Institute of Science and Technology, Campus UAB, 08193 Bellaterra (Barcelona), Spain}
 \email{rvicente@ifae.es}
 
\maketitle
  
  \vspace{-0.45cm}
  
\begin{abstract} 
Linearised relativistic elasticity equations of motion are considered for a rod and a spinning ring encountering a gravitational wave. In the case of the rod, the equations reduce to a wave equation with appropriate boundary conditions. Using Fourier transforms, the resonant frequencies are found and an explicit distributional solution is given, both for a plus- and a cross-polarised gravitational wave. In the case of the spinning ring, the equations are coupled wave equations with periodic boundary conditions. Using a Fourier series expansion, the system of wave equations is recast as a family of ordinary differential equations for the Fourier coefficients, which are then solved via Fourier transforms. The resonant frequencies are found, including simple approximate expressions for slowly rotating rings, and an explicit distributional solution is obtained in the case of the non-spinning ring. Interestingly, it is possible to tune the resonant frequencies by adjusting the angular velocity of the spinning ring.
\end{abstract}

{\small
\tableofcontents
}

\vspace{-1.5cm}
 
 \section{Introduction}
 
\noindent  The aim of this article is to study the linearised relativistic elasticity equations of motion for a  rod and a spinning ring encountering a gravitational wave. 

The existence of gravitational waves was predicted by Einstein already in 1916 \cite{Einstein16, Einstein18}, a year after the introduction of the theory of general relativity. He showed that
the linearised weak-field equations corresponding to a matter source with a time-varying mass quadrupole moment admit wave solutions  that travel at the speed of light.
However, their conclusive detection had to wait a century, with the observation in 2015 of the gravitational waves arising from a binary black hole merger  by the 
Laser Interferometer Gravitational-Wave Observatory (LIGO) (see \cite{LIGO}), ushering in a new era in astronomy. 

The detection of gravitational waves can be accomplished (at least conceptually) by monitoring the trajectories of free-falling test particles, given by timelike geodesics. To model a finite size detector, however, one should use the theory of relativistic elasticity, as it offers a coherent framework within general relativity which also accounts for the inevitable deformations that any real object is subject to (for background on the modern formulation of relativistic elasticity,  we refer the reader to \cite{BS03, Beig23} and the references therein). This theory has been used extensively to model extended astrophysical objects (see e.g. \cite{KS03, KS04, ABS09, BS08, AOS16, BS17, ANPR22a, ANPR22b, ANPR24a, ANPR24b}); the specific case of the response of elastic bodies to a gravitational wave has been recently considered in \cite{HMP, BCS, BMM}.

In this paper, we discuss relativistic elastic rods (open strings) and rings (closed strings), that is, one-dimensional elastic bodies whose internal energy depends only on their stretching, first studied by Carter \cite{Carter89, Carter89b} as models for superconducting cosmic strings (see also \cite{Natario14,  NQV} and references therein). We determine the linearised equations of motion in spacetimes modelling both plus- and cross-polarised gravitational waves.  
  
 In the case of a rod, we show that the equations reduce to a wave equation with appropriate boundary conditions. Using Fourier transforms, we find the resonant frequencies and give an explicit distributional solution. In the case of the spinning ring, we show that the equations written in cylindrical coordinates are coupled wave equations with periodic boundary conditions. Using a Fourier series expansion, we recast the system of wave equations as a family of ordinary differential equations for the Fourier coefficients, which we then solve via Fourier transforms. We find the resonant frequencies, including simple approximate expressions for slowly rotating rings, and obtain an explicit solution in the case of the non-spinning ring. Interestingly, we show that it is possible to tune the resonant frequencies by adjusting the angular velocity of the spinning ring.  This extends to general elastic laws the results obtained in \cite{Bol} for spinning Cosserat strings. 

The organisation of the paper is as follows. In \S\ref{section_preliminaries}, we briefly review the theory of relativistic elasticity of strings (that is, one-dimensional objects). In \S\ref{section_rod_plus}, we consider the case of a rod encountering a plus-polarised gravitational wave, and in \S\ref{section_rod_cross} we repeat this analysis for a cross-polarised gravitational wave. In \S\ref{section_ring}, we discuss the case of a spinning ring encountering a gravitational wave (for an arbitrary polarisation, since, due to the ring's symmetry, both polarisations have similar effects). Finally, we summarise and discuss our results in \S\ref{conclusion}.

We follow the conventions of \cite{MTW73, W84}, including a geometrised system of units, for which $c=G=1$. Greek letters $\mu, \nu, \ldots$ represent spacetime indices, running from $0$ to $3$, whereas capital Latin letters $A, B, \ldots$ represent indices in the string's worldsheet, taking the values $0$ and $1$. We used {\sc Mathematica} for symbolic and numerical computations, and also to produce Figure~\ref{fig:roots}.
 
 \section{Preliminaries} 
 \label{section_preliminaries}
 
\noindent  In this section we fix some notation, and recall the set-up from \cite{NQV}. 
 
We model an elastic string (that is, a one-dimensional elastic body whose internal energy depends only on its stretching) moving on a $4$-dimensional spacetime $(M,\bbg)$ by an embedding $X:\mR \times I \to M$, where $I \subset \mR$ is an interval labelling the points of the string. In the case of string loops (rings) we identify the endpoints of $I$ to obtain an embedding $X:\mR \times S^1 \to M$. The curve $\tau \mapsto X(\tau,\lambda)$ is the worldline of the point of the string labelled by $\lambda \in I$. For simplicity, we assume that the parameter $\lambda \in I$ is the arclength in the string's unstretched configuration.  
The embedding $X$ induces a metric
\begin{equation}\label{metric}
\bbh_{AB} = \bbg_{\mu\nu}(X) \partial_A X^\mu \partial_B X^\nu
\end{equation}
on $\mR \times I$, and we  identify $\mR \times I$ with its image $\Sigma = X(\mR \times I)$ (sometimes called the string's {\normalfont \it worldsheet}). If we choose a local orthonormal frame 
$\{\bbE_\tau, \bbE_\lambda\}$  
tangent to $\Sigma$ such that $\bbE_\tau$ is the $4$-velocity of the string's particles, then $ \frac{\partial X}{\partial \tau} = \alpha \bbE_\tau $ and
$ \frac{\partial X}{\partial \lambda} = \beta \bbE_\tau + \sigma \bbE_\lambda$
for some smooth local functions $\alpha, \beta, \sigma$. Note that $|\sigma|$ represents the factor by which the string is stretched according to an observer comoving with it, since $\bbE_\lambda$ gives the direction of simultaneity for such an observer. The components of the induced metric are then

\vspace{0.1cm}

$\phantom{AAAAAAAAAAaa}
[\bbh_{AB}] =
{\scaleobj{0.81}{ 
\left[
\begin{matrix}
-\alpha^2 & -\alpha \beta \\
- \alpha \beta & - \beta^2 + \sigma^2
\end{matrix}
\right]}} \, ,
$

\vspace{0.1cm}

\noindent 
and so
 $
h \equiv \det[\bbh_{AB}] = - \alpha^2 \sigma^2 = \bbh_{\tau \tau} \sigma^2.
$ 
Defining the {\normalfont \it number density} $n = \frac{1}{|\sigma|}$, we then have
\begin{equation} 
\label{n^2}
\textstyle 
n^2 = {\scaleobj{0.81}{\qfrac{\bbh_{\tau\tau}}{h}}}\, .
\end{equation}
To obtain the string's equations of motion we must choose an action

\vspace{0.1cm}

$\phantom{AAAAAAAAAAAa}
S = {\scaleobj{0.81}{\displaystyle \int_{\mR \times I}}} \calL(X,\partial X) \, d\tau d\lambda \, .
$

\vspace{0.1cm}

\noindent For an elastic string whose internal energy density $\rho$ depends only on its stretching, $\rho = F(n^2)$, the Lagrangian density is 
 $
\calL = F(n^2) \sqrt{-h} \, ,
$ 
where $h \equiv \det[\bbh_{AB}]$ and $n^2$ are given as functions of $(X, \partial X)$ from equations \eqref{metric} and \eqref{n^2}. This Lagrangian density reduces to the usual Newtonian Lagrangian density for an elastic string in the appropriate limit.

The equations of motion are obtained by computing the variation $\delta \calL$ of the Lagrangian density resulting from a variation $\delta X$ of the embedding. We define the string's energy momentum tensor $\bbT^{AB}$  by the relation
 $
\delta \calL = - \frac12 \sqrt{-h} \; \bbT^{AB}\; \delta \bbh_{AB} \, .
$ 
It can be shown (see \cite{NQV}) that

\vspace{0.1cm}

$\phantom{AAaa}
\bbT^{AB} = 2 n^2 F'(n^2) \bbU^A \bbU^B + (2 n^2 F'(n^2) - F(n^2))\bbh^{AB}\,\,,
$

\vspace{0.1cm}

\noindent 
where $\bbU^A$, $A=0,1$, denote the induced components on the string's worldsheet
 of 
the four-velocity $\bbE_\tau$ of the string's particles. Therefore, the string's energy density $\rho$ and the string's pressure $p$ are given by
\begin{equation} \label{pressure}
\rho = F(n^2) \,\, , \qquad p = 2 n^2 F'(n^2) - F(n^2) \,\, .
\end{equation}
The equations of motion are given by (see \cite{NQV})
\begin{equation} \label{EoM}
{\scaleobj{0.9}{\qfrac{1}{\sqrt{-h}}}}\partial_B \pmb{(} \sqrt{-h} \, \bbT^{AB}\partial_A X^\alpha \pmb{)} + \bbT^{AB} \Gamma^\alpha_{\mu\nu} \partial_A X^\mu \partial_B X^\nu = 0\,\,.
\end{equation}

The speeds of local perturbations travelling on a string can be obtained by linearising the equations of motion about a (possibly stretched) stationary string in Minkowski spacetime, aligned, say, with the $x$-axis. This corresponds to taking terms up to quadratic order in the Lagrangian obtained from the embedding
$$
{\scaleobj{0.96}{
\begin{array}{rcl}
t(\tau, \lambda) \!\!\!\!&=&\!\!\! \tau \, , \\[0.05cm]
x(\tau, \lambda) \!\!\!\!&=&\!\!\! {n_0}^{-1} \lambda + \delta x(\tau, \lambda) \, , \\[0.05cm]
y(\tau, \lambda) \!\!\!\!&=&\!\!\! \delta y(\tau, \lambda) \, , \\[0.05cm]
z(\tau, \lambda) \!\!\!\!&=&\!\!\! \delta z(\tau, \lambda) \, .
\end{array} }}
$$
Approximating $\bbh_{00}$, $h$, $F$  to  quadratic order, one obtains
$$
\begin{array}{rcl}
\calL = F(n^2) \sqrt{-h} \!\!\!\!&=&
F'({n_0}^2) {n_0}^2\;\! \pmb{\big(} \big( 2{n_0}^4 {\scaleobj{0.9}{\qfrac{F''({n_0}^2)}{F'({n_0}^2)}}} + {n_0}^2 \big) {\delta x'}^2 - \delta \dot{x}^2 \;\! \pmb{\big)}  \\[0.27cm]
\!\!\!\!&& \!\!\!+\;\! {\scaleobj{0.9}{\qfrac{1}{2}}} F({n_0}^2) {n_0}^2 \;\! \pmb{\big(} \big( {n_0}^2 - 2{n_0}^4 {\scaleobj{0.9}{\qfrac{F'({n_0}^2)}{F({n_0}^2)}}} \big) {\delta y'}^2 - \delta \dot{y}^2\;\! \pmb{\big)} \\[0.27cm]
\!\!\!\!&&\!\!\! + \;\!{\scaleobj{0.9}{\qfrac{1}{2}}} F({n_0}^2) {n_0}^2 \;\! \pmb{\big(} \big( {n_0}^2 - 2{n_0}^4 {\scaleobj{0.9}{\qfrac{F'({n_0}^2)}{F({n_0}^2)}}} \big) {\delta z'}^2 - \delta \dot{z}^2\;\!  \pmb{\big)} \, .
\end{array}
$$
So $\delta x$ satisfies the wave equation in the coordinates $(\tau,\lambda)$ with wave speed
$$
c' = n_0 \;\!\sqrt{2{n_0}^2 \;\!{\scaleobj{0.9}{\qfrac{F''({n_0}^2)}{F'({n_0}^2)} }}+ 1} \, ,
$$
whereas $\delta y$ and $\delta z$ satisfy the wave equation with wave speed
$$
s' = n_0\;\! \sqrt{1 - 2{n_0}^2\;\! {\scaleobj{0.9}{\qfrac{F'({n_0}^2)}{F({n_0}^2)}}}} \, .
$$
Since $\lambda = n_0 x$ for the stretched string, we see that the physical speed of sound for longitudinal waves is
\begin{equation}
\label{c^2}
c = \sqrt{2n^2\;\! {\scaleobj{0.9}{\qfrac{F''(n^2)}{F'(n^2)}}} + 1} 
= \sqrt{{\scaleobj{0.9}{\qfrac{dp}{d\rho}}}}\,\,,
\end{equation}
the same expression as the speed of sound for a perfect fluid, whereas the speed of sound for transverse waves is given by
$$
s = \sqrt{1 - 2n^2\;\! {\scaleobj{0.9}{\qfrac{F'(n^2)}{F(n^2)}}}} 
= \sqrt{{\scaleobj{0.9}{-\qfrac{p}{\rho}}}}\,\,,
$$
generalising the well-known classical result. 
A necessary condition for the stability of the stretched string is that $c$ and $s$ be real (otherwise there would exist exponentially growing modes in the limit of small wavelengths), that is, $\frac{dp}{d\rho}\geq 0$ and $p \leq 0$.

There are many possible choices for the  `elastic law' 
$
\rho = F(n^2)$, each corresponding to a different kind of elastic string. Some important examples (for a given constant energy density $\rho_0>0$ of the unstretched string) 
are the following:

\vspace{0.15cm}

\noindent 
{\bf Non-prestressed strings with constant longitudinal speed of sound ${\bm c\bm \geq \bm 0}$:}  Here $
\rho = {\scaleobj{0.72}{\qfrac{\rho_0}{c^2+1}}} (n^{c^2+1} \!+\! c^2)$, yielding $p \!=\! {\scaleobj{0.72}{\qfrac{\rho_0 c^2}{c^2+1}}} (n^{c^2+1} \!-\! 1)$. 
For $c=1$ we obtain the  `rigid' string, and for $c=0$ we have an incoherent dust string.

\vspace{0.15cm}

\noindent {\bf Strings with constant transverse speed of sound ${\bm s} \;\!{\bm \geq}\;\!{\bm  0}$:} This corresponds to  
$\rho = \rho_0 n^{1-s^2}$, giving $ p = - s^2 \rho$.
 For $s=1$ we obtain the Nambu-Goto string, and for $s=0$ we again have a dust string.

\vspace{0.15cm}

\noindent {\bf `Warm' cosmic string model with mass parameter ${\bm m }\;\!{\bm \geq}\;\!{\bm  0}$:}  Here 
$\rho = \sqrt{({\rho_0}^2 - m^4)n^2 + m^4}$, implying $ p = - \frac{m^4}{\rho}$ 
 (with $m^2 < \rho_0$). In this case, the longitudinal and transverse speeds of sound coincide. For $m=0$ we again have a dust string.

\vspace{0.15cm}

\noindent Depending on the elastic law, the string may have different properties, and we elaborate on these below. 

\vspace{0.15cm}

\noindent {\bf Existence of a relaxed configuration:}  If the pressure is zero when the string is not stretched nor compressed (that is, if the string is not pre-stressed), then $F$ must satisfy
$2F'(1) = F(1)$. Of the three models above, only the first satisfies this condition.

\vspace{0.15cm}

\noindent {\bf Weak energy condition:} The weak energy condition $\rho \geq 0$ and $\rho + p \geq 0$ is equivalent to $F(n^2) \geq 0$ and $ F'(n^2) \geq 0$. In particular, if the string satisfies the weak energy condition, then $\rho$ is a nondecreasing function of $n^2$. All the models above satisfy this condition.

\vspace{0.15cm}

\noindent {\bf $\;\!$Dominant $\;\!$energy $\;\!$condition$\;\!$:$\;\!$} The $\;\!$dominant $\;\!$energy $\;\!$condition $\rho \geq p \geq -\rho$ is equivalent to $F(n^2) \geq n^2 F'(n^2) \geq 0$. If the string satisfies the dominant energy condition, then it also satisfies the weak energy condition. Of the three models above, only the first two satisfy the dominant energy condition, and only for $c \leq 1$ and $s \leq 1$. (It is clear that if an elastic string satisfies the dominant energy condition, then its transverse speed of sound cannot exceed the speed of light.) 

\vspace{0.15cm}

\noindent {\bf Well-defined longitudinal speed of sound:}  If the longitudinal speed of sound is well defined, then from \eqref{c^2} we must have $F'(n^2)\neq 0$ and $\frac{dp}{d\rho} \geq 0$. Of the three models above, only the first and the third satisfy this condition.  (Technically, the second model also satisfies this condition in the trivial case $s=0$.) If the string also satisfies the weak energy condition, then $\rho$ is a strictly increasing function of $n^2$, and hence $p$ is a nondecreasing function of $n^2$.

 \section{Rod encountering a plus-polarised gravitational wave}
 \label{section_rod_plus}
 
 \subsection{Plus-polarised gravitational wave} 
 We now assume that $(M,\bbg)$ is a $4$-dimensional  spacetime  modelling a plus-polarised gravitational wave, 
 with a metric of the form 
 $$
 ds^2 =-dt^2 +(1+\varphi(t-z))\;\! dx^2 +(1-\varphi(t-z))\;\! dy^2 +dz^2 \, ,
 $$
 in a Cartesian coordinate chart. Thus the gravitational wave disturbance propagates along the $z$-direction of the chart, with the profile of the wave described by the smooth function $\varphi$. Furthermore, $\varphi$ is thought of as being small ($|\varphi| \ll 1$), 
 so that we can think of $(M,\bbg)$ as a perturbation of Minkowski spacetime solving the linearised Einstein equations (that is, solving the Einstein equations to first order in $\varphi$). 
 The nonzero connection coefficients for the Levi-Civita connection $\nabla_\bbg$ induced by $\bbg$ are given (up to first order) as follows (here and  henceforth we denote first order approximations by $\approx$):  
 $$
 \begin{array}{rcl}
   \Gamma^t_{xx}=-\Gamma^t_{yy}\!\!\!\!&\approx&\!\!\!\!
   {\scaleobj{0.9}{\qfrac{\varphi'(t-z)}{2}}} \, , \\
 \Gamma^x_{tx}=\Gamma^x_{xt}=-\Gamma^x_{xz}=-\Gamma^x_{zx}\!\!\!\!&\approx&\!\!\!\!{\scaleobj{0.9}{\qfrac{\varphi'(t-z)}{2}}} \, , \\
-\Gamma^y_{ty}=-\Gamma^y_{yt}=\Gamma^y_{yz}=\Gamma^y_{zy}\!\!\!\!&\approx&\!\!\!\!{\scaleobj{0.9}{\qfrac{\varphi'(t-z)}{2}}} \, , \\
 \Gamma^z_{xx}=-\Gamma^z_{yy}\!\!\!\!&\approx&\!\!\!\!{\scaleobj{0.9}{\qfrac{\varphi'(t-z)}{2}}} \, .
 \end{array}
 $$

 \subsection{Induced metric on the rod's worldsheet} 
 As explained in Section~\ref{section_preliminaries}, we model a rod in the $4$-dimensional spacetime $(M,\bbg)$ by an embedding $X:\mR\times I\rightarrow M$, where $I:=[0,L]\subset \mR$ is the interval labelling the points along the rod.  We use Cartesian coordinates, and assume that the rod is initially lying along the $x$-axis. Thus we have
 $$
 X(\tau,\lambda)=
 {\scaleobj{0.81}{
 \left[\begin{array}{c} t(\tau,\lambda)\\[0.06cm]
 x(\tau,\lambda)\\[0.06cm]
 y(\tau,\lambda)\\[0.06cm]
 z(\tau,\lambda)\end{array}\right]}}
 =
   {\scaleobj{0.81}{\left[\begin{array}{c} \tau \\[0.06cm] \lambda+\xi(\tau,\lambda)\\[0.06cm] \eta(\tau,\lambda)\\[0.06cm] \zeta(\tau,\lambda)
  \end{array}\right]}} \, ,
  $$
  where $(\tau,\lambda)\mapsto \xi(\tau,\lambda),\;\eta(\tau,\lambda), \;\zeta(\tau,\lambda)$ describe the small perturbations of the coordinates of the particles along the rod.   The metric $\bbh=X^* \bbg$ on $\Sigma:=X(\mR\times I)$   has the components 
  $$
  \bbh_{AB}=\bbg_{\mu\nu}(X) \partial_A X^\mu \;\! \partial_BX^\nu \, , \quad A,B\in \{\lambda,\tau\} \, ,
  $$
   given below (blank entries in matrices are $0$):
  $$
  \bbh_{\tau\tau}\!=\!
   {\scaleobj{0.81}{\left[\!\begin{array}{cccc} \;1 \; &\;\partial_\tau \xi\; &\;\partial_\tau \eta\; & \;\partial_\tau\zeta\; \end{array}\!\right] 
  \left[\!\begin{array}{cccc} -1 & & & \\ & 1+\varphi(t-z) && \\ &&1-\varphi (t-z) & \\ &&&1\end{array}\!\right]\!\left[\!\begin{array}{c} 1 \\[0.06cm] \partial_\tau\xi \\[0.06cm] \partial_\tau\eta \\[0.06cm] \partial_\tau\zeta \end{array}\!\right]}}
 \approx -1 \, ,
  $$
  and similarly 
  $ \bbh_{\tau \lambda}=\bbh_{\lambda\tau} \approx  \partial_\tau\xi$ and 
  $\bbh_{\lambda \lambda}\approx 1+2\partial_\lambda \xi +\varphi(t-z)$.
  Therefore,
  $$
  [\bbh_{AB}]\approx
  {\scaleobj{0.81}{\left[\begin{array}{cc} -1\;\; & \;\;\partial_\tau\xi\\[0.06cm]\partial_\tau\xi\;\;&\;\;1+2\partial_\lambda \xi +\varphi(t-z)\end{array}\right]}} \, ,
  $$
  with determinant
  $$
  h=\det [\bbh_{AB}]\approx -1-2\partial_\lambda \xi -\varphi(t-z) \, .
  $$ 
  Also, we have 
  \begin{eqnarray*}
  [\bbh^{AB}]\!\!\!\!&:=&\!\!\!\![\bbh_{AB}]^{-1}\!\approx 
  {\scaleobj{0.81}{\left[\!\begin{array}{cc} -1\; & \;\partial_\tau\xi\\[0.06cm] \partial_\tau\xi\;&1\!+\!2\partial_\lambda \xi \!+\!\varphi(t\!-\!z)\end{array}\!\right]^{-1}}}
  \!\!\!=\!{\scaleobj{0.72}{\qfrac{1}{h}}}
   {\scaleobj{0.81}{\left[\!\begin{array}{cc} 1\!+\!2\partial_\lambda \xi \!+\!\varphi(t-z) &- \partial_\tau\xi\\[0.06cm]-\partial_\tau\xi\;&\;-1\end{array}\!\right]}}
  \\
  \!\!\!&\approx&\!\!\!\! 
  {\scaleobj{0.81}{\left[\!\begin{array}{cc} -1 \;&\; \partial_\tau\xi\\[0.06cm] \partial_\tau\xi \;&\; 1-2\partial_\lambda \xi -\varphi(t-z) \end{array}\right]}} \, .
  \end{eqnarray*}
  We obtain the number density $n$ as
  $$
  n^2={\scaleobj{0.9}{\qfrac{\bbh_{\tau\tau}}{h}}}
  \approx {\scaleobj{0.9}{\qfrac{-1}{-1-2\partial_\lambda \xi -\varphi(t-z)}}}
  \approx 1-2 \partial_\lambda \xi -\varphi(t-z)=1+\Delta \, ,
  $$
  where 
  $$
  \Delta:=-2 \partial_\lambda \xi -\varphi(t-z) \, . 
  $$
  If we consider a non-prestressed rod with a given elastic law $\rho = F(n^2)$, then we have
  $$
  F(1)=\rho_0 \, , \qquad F'(1) = {\scaleobj{0.9}{\qfrac{\rho_0}{2}}} \, ,
  $$
  where $\rho_0$ is the density of the relaxed configuration. Moreover, we see from \eqref{c^2} that
  $$
  F''(1) = {\scaleobj{0.9}{\qfrac{\rho_0 (c^2-1)}{4}}} \, ,
  $$
  where $c\geq 0$ is the longitudinal speed of sound in the undeformed state. From the first-order Taylor expansions around $n^2=1$ of \eqref{pressure}, we get 
  $$
  \rho \approx \rho_0 \big( 1 + {\scaleobj{0.9}{\qfrac{\Delta}{2}}}\big) \, , \qquad
  p \approx \rho_0 c^2 {\scaleobj{0.9}{\qfrac{\Delta}{2}}} \, .
  $$
  If $\bbU$ denotes the four-velocity of the particles of the rod, then 
   $$
  \bbU={\scaleobj{0.9}{\qfrac{1}{\sqrt{-\bbh_{\tau\tau}}} }}\partial_\tau X \, ,
  $$ 
  and so $
  [\bbU^A]=\left[\begin{smallmatrix} 1\\[0.06cm] 0\end{smallmatrix}\right] \, .
  $ 
  The energy momentum tensor $\bbT$ has the components $\bbT^{AB}$ given by 
  \begin{eqnarray*}
  [\bbT^{AB}]\!\!\!&=&\!\!\!\!
  (\rho+p)\;\![\bbU^A \bbU^B]+p\;\![\bbh^{AB}]
  \\
  \!\!\!&\approx &\!\!\!\!
  (\rho+p) 
  {\scaleobj{0.81}{\left[\!\begin{array}{cc} 1 \;&\; 0\\[0.06cm] 0 \;&\; 0\end{array}\right]}}
  +
  p 
  {\scaleobj{0.81}{\left[\!\begin{array}{cc} -1 \;&\; \partial_\tau\xi\\[0.06cm] \partial_\tau\xi \;&\; 1-2\partial_\lambda \xi -\varphi(t-z) \end{array}\right]}}
 \approx 
 \rho_0 {\scaleobj{0.81}{\Bigg[\!\begin{array}{cc} 1+
 {\scaleobj{0.9}{\qfrac{\Delta}{2}}}\; &\; 0 \\ 0 \;&\; {\scaleobj{0.9}{\qfrac{c^2\Delta}{2}}}\end{array}\Bigg]}} \, .
  \end{eqnarray*}
 The $x$-component of the equations of motion \eqref{EoM} is then given by
 $$
 \begin{array}{rl}
0=&\!\!\! {\scaleobj{0.9}{\qfrac{1}{\sqrt{-h}}}} \;\! \partial_\tau \pmb{(}\sqrt{-h} \;\!\bbT^{\tau \tau}  \;\!\partial_\tau x\pmb{)}
  +
  {\scaleobj{0.9}{\qfrac{1}{\sqrt{-h}}}}  \;\!\partial_\lambda \pmb{(}\sqrt{-h}  \;\!\bbT^{\lambda \lambda} \;\! \partial_\lambda x\pmb{)}\\[0.33cm]
 & +\;\!2 \;\! \bbT^{\tau \tau}  \Gamma^{x}_{tx} ( \partial_\tau t )(  \partial_\tau x) 
  +2 \;\! \bbT^{\tau \tau}  \Gamma^{x}_{xz} ( \partial_\tau x )(  \partial_\tau z)\\[0.24cm]
  &+\;\!2  \;\!\bbT^{\lambda \lambda}  \Gamma^{x}_{tx} ( \partial_\lambda t )(  \partial_\lambda x) 
  +2  \;\!\bbT^{\lambda \lambda}  \Gamma^{x}_{xz} ( \partial_\lambda x )(  \partial_\lambda z) \, ,
 \end{array}
 $$
 that is,
 $$
  \begin{array}{rl}
 0\;\! \approx&\!\!\! {\scaleobj{0.81}{\qfrac{1}{\sqrt{1-\Delta}} }}\partial_\tau \pmb{\big(}\sqrt{1-\Delta} (1+{\scaleobj{0.81}{\qfrac{\Delta}{2}}}) \partial_\tau x\pmb{\big)}
  +
  {\scaleobj{0.81}{\qfrac{1}{\sqrt{1-\Delta}}}} \partial_\lambda \pmb{\big(}\sqrt{1-\Delta} {\scaleobj{0.81}{\qfrac{c^2\Delta}{2} }} \partial_\lambda x\pmb{\big)}\\[0.33cm]
  &+\;\! \big(1+{\scaleobj{0.81}{\qfrac{\Delta}{2}}}\big) \varphi'(t-z) ( 1 )(  \partial_\tau x) 
  - \big(1+{\scaleobj{0.81}{\qfrac{\Delta}{2}}}\big)  \varphi'(t-z) ( \partial_\tau x )(  \partial_\tau z)\\[0.27cm]
  &+\;\!  {\scaleobj{0.81}{\qfrac{c^2\Delta}{2}}} \varphi'(t-z) ( 0)(  \partial_\lambda x) 
  -  {\scaleobj{0.81}{\qfrac{c^2\Delta}{2}}}  \varphi'(t-z) ( \partial_\lambda x )(  \partial_\lambda z) \, .
 \end{array}
 $$
 Thus we obtain
 $$
 \partial_\tau^2 \xi +{\scaleobj{0.9}{\qfrac{c^2}{2}}}\partial_\lambda \pmb{(}-2 \partial_\lambda \xi -\varphi(t-z)\pmb{)}(1+\partial_\lambda \xi)\approx 0 \, ,
 $$
which yields the usual wave equation 
 $$
 \partial_\tau^2 \xi -c^2 \partial_\lambda^2 \xi\approx 0 \, .
 $$
In a similar manner, one can also derive that 
$$
\partial_\tau^2 \eta\;\!\approx \;\!0\;\;\textrm{ and } \;\;\partial_\tau^2 \zeta\;\!\approx \;\!0 \, .
$$
These last two equations are trivial, and correspond to inertial motion of the whole rod along the $y$ or the $z$-axis. Therefore, we will only analyse the equation for $\xi$.

\subsection{The boundary value problem for $\xi$ and for $\partial_\lambda \xi$} 

The boundary conditions are obtained by setting $p=0$ (that is, $\Delta=0$) at the endpoints $\lambda=0$ and $\lambda=L$. This results in the conditions 
$$
\left. \begin{array}{rcl}
(\partial_\lambda \xi)(\tau, 0)\!\!\!&=&\!\!\!\!-\frac{1}{2} \varphi (\tau)\\[0.21cm]
(\partial_\lambda \xi)(\tau, L)\!\!\!&=&\!\!\!\!-\frac{1}{2} \varphi (\tau)
\end{array}\right\} \textrm{ for all }\tau\in \mR \, .
$$
It is clear that if we assume that $\mR\times [0,L]\mapsto \xi(\tau,\lambda)$ is a solution to the wave equation $\partial_\tau^2 \xi -c^2 \partial_\lambda^2 \xi=0$ with the above  boundary conditions, then $\partial_\lambda\xi$ satisfies the following boundary value problem:
$$
\begin{array}{lll}
\textrm{\bf (PDE)} &  
\partial_\tau^2 (\partial_\lambda \xi) -c^2 \partial_\lambda^2 
(\partial_\lambda \xi)=0 & (\tau \in \mR, \;\lambda \in [0,L] ) \, ,
\\[0.24cm]
\textrm{\bf (BC)} & 
\left\{ \begin{array}{rcl}
(\partial_\lambda \xi)(\tau, 0)\!\!\!&=&\!\!\!\!-\frac{1}{2} \varphi (\tau)\\[0.15cm]
(\partial_\lambda \xi)(\tau, L)\!\!\!&=&\!\!\!\!-\frac{1}{2} \varphi (\tau)
\end{array}\right\} & (\tau\in \mR) \, .
\end{array}
$$
We use the method of Fourier transforms, and set 
$$
(\widehat{\partial_\lambda \xi})(\omega,\lambda)
:={\scaleobj{0.9}{\displaystyle \int_{-\infty}^\infty}} (\partial_\lambda \xi)(\tau,\lambda)e^{-i\omega \tau} d\tau \quad \;\;(\omega\in \mR, \lambda \in[0,L]) \, .
$$
Then 
\begin{eqnarray*}
-\omega^2 (\widehat{\partial_\lambda \xi})(\omega, \lambda)\!\!\!&=&\!\!\!\! 
{\scaleobj{0.9}{\displaystyle\int_{-\infty}^\infty}} \partial_\tau^2 (\partial_\lambda \xi)(\tau,\lambda)e^{-i\omega \tau} d\tau \, ,\\
-c^2 \partial_\lambda^2 (\widehat{\partial_\lambda \xi})(\omega \lambda)
\!\!\!&=&\!\!\!\! 
{\scaleobj{0.9}{\displaystyle\int_{-\infty}^\infty}}
 -c^2 \partial_\lambda^2 (\partial_\lambda \xi)(\tau,\lambda)e^{-i\omega \tau} d\tau \, .
\end{eqnarray*}
Adding these, we get 
$$
\partial_\lambda^2 (\widehat{\partial_\lambda \xi})(\omega, \lambda)
+{\scaleobj{0.9}{\qfrac{\omega^2}{c^2}}} (\widehat{\partial_\lambda \xi})(\omega, \lambda)=0 \, .
$$
The general solution is given by 
$$
(\widehat{\partial_\lambda \xi})(\omega, \lambda)
=
A(\omega) \cos \big({\scaleobj{0.9}{\qfrac{\omega}{c}}}\lambda\big)
+B(\omega) \sin \big({\scaleobj{0.9}{\qfrac{\omega}{c}}}\lambda\big)
$$
for some maps $\mR\owns \omega \mapsto A(\omega),B(\omega)\in \mR$. 
Let  $\widehat{\varphi}$ denote the Fourier transform for $\varphi$. Then 
the boundary conditions {\bf (BC)} give 
$$
\begin{array}{l}
-{\scaleobj{0.9}{\qfrac{\widehat{\varphi}(\omega)}{2}}}
=(\widehat{\partial_\lambda \xi})(\omega, 0)
=A(\omega) \cos 0+B(\omega) \sin 0=A(\omega) \, ,\\[0.27cm]
-{\scaleobj{0.9}{\qfrac{\widehat{\varphi}(\omega)}{2}}}
=(\widehat{\partial_\lambda \xi})(\omega, L)
= -{\scaleobj{0.9}{\qfrac{\widehat{\varphi}(\omega)}{2}}}
\cos \big({\scaleobj{0.9}{\qfrac{\omega}{c}}}L\big)
+B(\omega) \sin \big({\scaleobj{0.9}{\qfrac{\omega}{c}}}L\big) \, .
\end{array}
$$
Since $B(\omega)$ is a tempered distribution, the solution of the second equation is
$$
\textstyle 
B(\omega)=-{\scaleobj{0.9}{\qfrac{\widehat{\varphi}(\omega)(1-\cos ({\scaleobj{0.72}{\qfrac{\omega}{c}}}L)) }{2\sin ({\scaleobj{0.72}{\qfrac{\omega}{c}}}L)}}}
 + \sum\limits_{m\in\mathbb{Z}} c_m \delta_{{\scaleobj{0.6}{\qfrac{m\pi c}{L}}}}(\omega) \, ,
$$
as the zeroes $\omega = {\scaleobj{0.72}{\qfrac{m\pi c}{L}}}$ of the function $\sin ({\scaleobj{0.72}{\qfrac{\omega}{c}}}L)$ are simple.
The infinite sum above is the Fourier transform of the function
$$
\textstyle 
{\scaleobj{0.9}{
 \qfrac{1}{2\pi} \displaystyle \int_{-\infty}^{\infty}}}
  \Big(  \sum\limits_{m\in\mathbb{Z}} c_m \delta_{{\scaleobj{0.6}{\qfrac{m\pi c}{L}}}}(\omega) \Big) \sin \big({\scaleobj{0.9}{\qfrac{\omega}{c}}}\lambda\big) 
  e^{i\omega \tau} d\omega
=  {\scaleobj{0.9}{\qfrac{1}{2\pi}}} \sum\limits_{m\in\mathbb{Z}} c_m 
\sin \big({\scaleobj{0.9}{\qfrac{m\pi}{L}}}\lambda\big) e^{i{\scaleobj{0.6}{\qfrac{m\pi c}{L}}} \tau},
$$
which is the general solution of the wave equation with homogeneous boundary conditions written in Fourier series form. We assume that the rod is initially at rest, i.e., that the motion of the rod occurs purely in response to the gravitational wave perturbation, and so $c_m=0$ for all $m\in\mathbb{Z}$. Hence 
\begin{align*}
(\widehat{\partial_\lambda \xi})(\omega, \lambda)
& =
- {\scaleobj{0.9}{\qfrac{\widehat{\varphi}(\omega)}{2}}}
\cos \big( {\scaleobj{0.9}{\qfrac{\omega}{c}}}\lambda\big)
\!-\! {\scaleobj{0.9}{\qfrac{\widehat{\varphi}(\omega)(1-\cos ( {\scaleobj{0.72}{\qfrac{\omega}{c}}}L)) }{2\sin ( {\scaleobj{0.72}{\qfrac{\omega}{c}}}L)}}}\sin \big( {\scaleobj{0.9}{\qfrac{\omega}{c}}}\lambda\big) \\
& =
- {\scaleobj{0.9}{\qfrac{\widehat{\varphi}(\omega)}{2}}}
  {\scaleobj{0.9}{\qfrac{\cos ( {\scaleobj{0.72}{\qfrac{\omega}{c}}}
  ( {\scaleobj{0.72}{\qfrac{L}{2}}} -\lambda))}{\cos ( {\scaleobj{0.72}{\qfrac{\omega}{c}}} 
   {\scaleobj{0.72}{\qfrac{L}{2}}})}}} \, .
\end{align*}
The resonant frequencies of the rod, where the response to the gravitational wave signal will be stronger, are then given by
$$
\omega = {\scaleobj{0.9}{\qfrac{c}{L}}} (\pi + 2m\pi), \qquad m\in\mathbb{Z}\, .
$$
 We show below that the above `frequency-domain' description corresponds to the following `time domain' description:
 \begin{equation}
 \label{15_2_23_1547}
 \partial_\lambda \xi=- {\scaleobj{0.9}{\qfrac{\varphi}{2}}}  \ast T \, , 
 \end{equation}
 where $\ast$ denotes convolution, and $T$ is the tempered distribution 
 \begin{equation}
 \label{15_2_23_1548}
 \textstyle 
 T:=\sum\limits_{m=0}^\infty (-1)^m\delta_{ {\scaleobj{0.6}{\qfrac{\lambda+mL}{c}}}}-\sum\limits_{m=1}^\infty (-1)^m \delta_{ {\scaleobj{0.6}{\qfrac{-\lambda +mL}{c}}}} \, .
 \end{equation}
 In other words, $T$ is the Fourier transform of the function
 $$
 f(\omega)={\scaleobj{0.9}{\qfrac{\cos ( {\scaleobj{0.72}{\qfrac{\omega}{c}}}
  ( {\scaleobj{0.72}{\qfrac{L}{2}}} -\lambda))}{\cos ( {\scaleobj{0.72}{\qfrac{\omega}{c}}} 
   {\scaleobj{0.72}{\qfrac{L}{2}}})}}}
 $$ 
(note that $f$ is not locally integrable, but it can be interpreted as a tempered distribution by taking the principal value of its integral 
against tempered  functions).

In what follows, we suppose that the smooth function $\varphi:\mR\rightarrow \mR$ is compactly supported. We use the notation $\calD(\mR)$ for the space of compactly supported smooth (i.e., infinitely differentiable) functions on $\mR$.  The support of a function $\psi:\mR\rightarrow \mC$ is denoted by $\textrm{supp}\;\! \psi$. For an open set $U\subset \mR$, $\calD(U)=\{\psi \in \calD(\mR): \textrm{supp}\;\! \psi\subset U\}$.  We denote the space of tempered test functions on $\mR$ by $\calS(\mR)$, and the space of tempered distributions by $\calS'(\mR)$. For preliminaries on distributions, we refer the reader to \cite{Sch}. 
 
 \begin{proposition}
 \begin{spacing}{1.3}
 Suppose that $\lambda \in (0,L)$. Let $T$ be the distribution given by 
 $T=\sum\limits_{m=0}^\infty (-1)^m\delta_{ {\scaleobj{0.6}{\qfrac{\lambda+mL}{c}}}}-\sum\limits_{m=1}^\infty (-1)^m \delta_{ {\scaleobj{0.6}{\qfrac{-\lambda +mL}{c}}}}.$ Then $T$ is tempered$,$ that is$,$ $T\in \calS'(\mR)$.
 \end{spacing}
 \vspace{-0.21cm}
 \end{proposition} 
 \begin{proof} Let $\psi \in \calS(\mR)$. In particular, 
 $
 \sup\limits_{x\in \mR} |x^2 \psi(x)|=:M<\infty.
 $ 
 Thus 
 $$
 \begin{array}{rcl}
 |\langle T, \psi \rangle|
 \!\!\!&\leq & \!\!\! \sum\limits_{m=0}^\infty 
\big|\psi\big( {\scaleobj{0.9}{\qfrac{\lambda+mL}{c}}}\big)\big|
 +\sum\limits_{m=1}^\infty 
 \big|\psi\big( {\scaleobj{0.9}{\qfrac{-\lambda+mL}{c}}}\big)\big|\\[0.42cm]
 \!\!\!&\leq & \!\!\! \sum\limits_{m=0}^\infty  
 {\scaleobj{0.9}{\qfrac{M}{( {\scaleobj{0.72}{\qfrac{\lambda+mL}{c}}})^2}}}
 +\sum\limits_{m=1}^\infty
  {\scaleobj{0.9}{\qfrac{M}{( {\scaleobj{0.72}{\qfrac{-\lambda+mL}{c}}})^2}}}
  \leq CM \, , 
 \end{array}
 $$
 where $C:=\sum\limits_{m=0}^\infty 
  {\scaleobj{0.9}{\qfrac{1}{( {\scaleobj{0.72}{\qfrac{\lambda+mL}{c}}})^2}}} 
  +\sum\limits_{m=1}^\infty
   {\scaleobj{0.9}{\qfrac{1}{( {\scaleobj{0.72}{\qfrac{-\lambda+mL}{c}}})^2}}}<\infty$. So $T\in \calS'(\mR)$. 
 \end{proof}

 \begin{proposition}
 \begin{spacing}{1.41}
 Let $\lambda \in (0,L)$. 
Suppose that the smooth function $\varphi :\mR\rightarrow\mR$ is compactly supported. 
 Consider the distribution $T\in \calS'(\mR)$ given by $T=\sum\limits_{m=0}^\infty (-1)^m\delta_{ {\scaleobj{0.6}{\qfrac{\lambda+mL}{c}}}}
 -\sum\limits_{m=1}^\infty (-1)^m \delta_{ {\scaleobj{0.6}{\qfrac{-\lambda +mL}{c}}}}.$ 
 Then $S:=- {\scaleobj{0.72}{\qfrac{\varphi}{2}}}  \ast T$ is a weak solution of $\partial_\tau^2S -c^2 \partial_\lambda^2S=0$ in $\mR \times (0,L)$  satisfying the boundary conditions 
 $S(\cdot, 0)=- {\scaleobj{0.72}{\qfrac{1}{2}}}\varphi=S(\cdot, L)$.
 \end{spacing}
  \vspace{-0.21cm}
 \end{proposition}
  \begin{proof}  \begin{spacing}{1.2}It is enough to consider the case when $T$ is just one of its summands, i.e., $T= \delta_{ {\scaleobj{0.51}{\qfrac{\pm \lambda+mL}{c}}}}$ for some integer $m$. Let $\psi \in \calD(\mR)$, and $\chi\in \calD( (0,L))$. Then we have  \end{spacing}
  
  \vspace{-0.6cm}
  
 $$
 \begin{array}{rcl}
 \langle S, \psi \otimes \chi\rangle 
 \!\!\!&=&\!\!\!
 {\scaleobj{0.9}{\displaystyle \int_0^L}} \big\langle - {\scaleobj{0.9}{\qfrac{\varphi}{2}}}  \ast T, \psi\big\rangle \;\!\chi(\lambda)\;\!d\lambda 
 \\[0.33cm]
 \!\!\!&=&\!\!\!
  {\scaleobj{0.9}{\displaystyle \int_0^L\!\!\int_{-\infty}^\infty}}\!\!\!
   - {\scaleobj{0.9}{\qfrac{\varphi(\tau)}{2}}} 
   \big\langle  \delta_{ {\scaleobj{0.6}{\qfrac{\pm \lambda+mL}{c}}}}, \psi(\tau+\cdot)\big\rangle \;\! d\tau \;\! \chi(\lambda)\;\!d\lambda 
 \\[0.33cm]
 \!\!\!&=&\!\!\!
 {\scaleobj{0.9}{\displaystyle \int_0^L\!\!\int_{-\infty}^\infty}}\!\!\!
  - {\scaleobj{0.9}{\qfrac{\varphi(\tau)}{2}}} 
  \psi \big(\tau+ {\scaleobj{0.9}{\qfrac{\pm \lambda+mL}{c}}}\big) \;\! d\tau \;\!\chi(\lambda)\;\!d\lambda \, .
 \end{array}
 $$
 So 
 $$ {\scaleobj{0.97}{
 \langle \partial_\tau^2  S, \psi \otimes \chi\rangle 
\!=\!
\langle S, \psi'' \!\otimes \chi\rangle 
\!=\!
 {\scaleobj{0.9}{\displaystyle \int_0^L\!\!\int_{-\infty}^\infty}}\!\!\!- {\scaleobj{0.9}{\qfrac{\varphi(\tau)}{2}}} \psi'' \big(\tau\!+\! {\scaleobj{0.9}{\qfrac{\pm \lambda\!+\!mL}{c}}}\big)  \chi(\lambda) d\tau d\lambda \, ,}}
$$ 
and 
$$
\!\!
{\scaleobj{0.97}{
\begin{array}{rcl}
\langle -c^2\partial_\lambda^2  S, \psi \otimes \chi\rangle 
\!\!\!\!&=&\!\!\!\!
 \langle S, \psi \otimes (-c^2)\chi''\rangle 
 \\[0.33cm]
\!\!\!\!&=&\!\!\!\!
  {\scaleobj{0.9}{\displaystyle \int_0^L\!\!\int_{-\infty}^\infty}} \!\!\!
  -{\scaleobj{0.9}{\qfrac{\varphi(\tau)}{2}}}
   \psi \big(\tau+{\scaleobj{0.9}{\qfrac{\pm \lambda+mL}{c}}}\big) 
    (-c^2)\;\!\chi''(\lambda) \;\!d\tau \;\!d\lambda 
 \\[0.33cm]
 \!\!\!\!&=&\!\!\!\!
 - {\scaleobj{0.9}{\displaystyle \int_0^L\!\!\int_{-\infty}^\infty}}\!\!\!
  -{\scaleobj{0.9}{\qfrac{\varphi(\tau)}{2}}}
   \big({\scaleobj{0.9}{\qfrac{\pm 1}{c}}}\big)
   \psi' \big(\tau\!+\!{\scaleobj{0.9}{\qfrac{\pm \lambda+mL}{c}}}\big)  
   (-c^2)\;\!\chi'(\lambda)\;\! d\tau\;\! d\lambda\\[0.33cm]
 \!\!\!\!&=&\!\!\!\!
  {\scaleobj{0.9}{\displaystyle \int_0^L\!\!\int_{-\infty}^\infty}} \!\!\!
  -{\scaleobj{0.9}{\qfrac{\varphi(\tau)}{2}}} 
  \big({\scaleobj{0.9}{\qfrac{1}{c^2}}}\big)
  \psi'' \big(\tau+{\scaleobj{0.9}{\qfrac{\pm \lambda+mL}{c}}}\big)  
  (-c^2)\;\!\chi(\lambda)\;\!d\tau \;\!d\lambda\\[0.33cm]
  \!\!\!\!&=&\!\!\!\!
  - {\scaleobj{0.9}{\displaystyle \int_0^L\!\!\int_{-\infty}^\infty}} \!\!\!
  -{\scaleobj{0.9}{\qfrac{\varphi(\tau)}{2}}} 
  \psi'' \big(\tau+{\scaleobj{0.9}{\qfrac{\pm \lambda+mL}{c}}}\big)  
  \;\!\chi(\lambda) \;\!d\tau\;\! d\lambda \, .
 \end{array}}}
 $$
 In the above we used integration by parts with respect to the $\lambda$ variable twice in order to get the equalities in the third and forth lines, and also the fact that $\chi$ vanishes at the endpoints since $\textrm{supp} \;\! \chi \subset (0,L)$. Thus for all $\psi \in \calD(\mR)$, and $\chi\in \calD( (0,L))$, we have $\langle \partial_\tau^2  S-c^2 \partial_\lambda^2 S, \psi \otimes \chi\rangle =0$. 
 By the linearity of $S$, and the density of $\calD(\mR)\otimes \calD((0,L))$ in $\calD(\mR\times (0,L))$, it follows that 
 $\partial_\tau^2  S-c^2\partial_\lambda^2 S=0$. 
 The given boundary conditions are satisfied, since 
 $$
 \textstyle 
 S(\cdot, 0)
 \!=\!
 -{\scaleobj{0.9}{\qfrac{\varphi}{2}}}
 \ast \pmb{\big(} \sum\limits_{m=0}^\infty 
 (-1)^m\delta_{{\scaleobj{0.6}{\qfrac{mL}{c}}}}
 -\sum\limits_{m=1}^\infty (-1)^m 
 \delta_{{\scaleobj{0.6}{\qfrac{mL}{c}}}}\pmb{\big)}
 \!=\!
 -{\scaleobj{0.9}{\qfrac{\varphi}{2}}}\ast ((-1)^0 \delta_0)
 \!=\!-{\scaleobj{0.9}{\qfrac{\varphi}{2}}}
 $$
 and 
 $$
 \begin{array}{rcl}
 S(\cdot, L)\!\!\!&=&\!\!\!
 -{\scaleobj{0.9}{\qfrac{\varphi}{2}}}\ast \pmb{\big(}\sum\limits_{m=0}^\infty (-1)^m\delta_{{\scaleobj{0.6}{\qfrac{L+mL}{c}}}}-\sum\limits_{m=1}^\infty (-1)^m \delta_{{\scaleobj{0.6}{\qfrac{-L +mL}{c}}}}\pmb{\big)}
 \\[0.2cm]
 \!\!\!&=&\!\!\!
 -{\scaleobj{0.9}{\qfrac{\varphi}{2}}}
 \ast \pmb{\big(} (\delta_{{\scaleobj{0.6}{\qfrac{L}{c}}}} -\delta_{{\scaleobj{0.6}{\qfrac{2L}{c}}}} +-\cdots) -(-\delta_0+\delta_{{\scaleobj{0.6}{\qfrac{L}{c}}}} -\delta_{{\scaleobj{0.6}{\qfrac{2L}{c}}}} +-\cdots)\pmb{)} 
  \\[0.3cm]
 \!\!\!&=&\!\!\!-{\scaleobj{0.9}{\qfrac{\varphi}{2}}}\ast \delta_0=-{\scaleobj{0.9}{\qfrac{\varphi}{2}}} \, .
  \end{array}
 $$
 
 \vspace{-0.6cm}
 
 \end{proof}
 
 \begin{remark}
 What can actually be measured along the rod is the stretching or the pressure (see \cite{RSBBHLF, SMP} for possible techniques to do so), both of which are proportional to
$$
\Delta = -2 \partial_\lambda \xi -\varphi(\tau) = \varphi \ast T - \varphi(\tau) \, .
$$
From this expression, it is clear that $\Delta$ has the same resonant frequencies as $\partial_\lambda \xi$. For completeness, the `time domain' expression for $\Delta$ is
$$
\textstyle
\Delta(\tau,\lambda) =\sum\limits_{m=0}^\infty (-1)^m \varphi\big(\tau - { {\scaleobj{0.9}{\qfrac{\lambda+mL}{c}}}}\big)-\sum\limits_{m=1}^\infty (-1)^m \varphi\big(\tau + { {\scaleobj{0.9}{\qfrac{\lambda -mL}{c}}}}\big) - \varphi(\tau) \, .
$$
\end{remark}
 
 \section{Rod encountering a cross-polarised gravitational wave}
 \label{section_rod_cross}
 
 \subsection{Cross-polarised gravitational wave} 
 
 Consider now a $4$-dimensional spacetime $(M,\bbg)$ modelling a cross-polarised gravitational wave, with metric 
 $$
 ds^2 =-dt^2 +dx^2+2\psi(t-z)dxdy+dy^2 +dz^2 \, ,
 $$
 in a Cartesian coordinate chart. Thus the gravitational wave disturbance propagates along the $z$-direction of the chart, with the profile of the wave described by the smooth function $\psi$. Furthermore, $\psi$ is thought of as small ($|\psi| \ll 1$), 
 so that we can think of $M$ as a perturbation of Minkowski spacetime solving the linearised Einstein equations (that is, solving the Einstein equations to first order in $\psi$). 
 The nonzero connection coefficients for the Levi-Civita connection $\nabla_\bbg$ induced by $\bbg$ are given to first order by:
$$ 
\begin{array}{rcl}
   \Gamma^t_{xy}=\Gamma^t_{yx}\!\!\!&\approx&
   \!\!\!\!-{\scaleobj{0.9}{\qfrac{\psi'(t-z)}{2}}} \, , \\
 \Gamma^x_{ty}=\Gamma^x_{yt}=-\Gamma^x_{yz}=-\Gamma^x_{zy}\!\!\!&\approx&
 \!\!\!\!-{\scaleobj{0.9}{\qfrac{\psi'(t-z)}{2}}} \, , \\
\Gamma^y_{tx}=\Gamma^y_{xt}=-\Gamma^y_{xz}=-\Gamma^y_{zx}\!\!\!&\approx&
\!\!\!\!-{\scaleobj{0.9}{\qfrac{\psi'(t-z)}{2}}} \, , \\
 \Gamma^z_{xy}=\Gamma^z_{yx}\!\!\!&\approx&\!\!\!\!
 -{\scaleobj{0.9}{\qfrac{\psi'(t-z)}{2}}} \, .
 \end{array}
 $$
 
  \subsection{Induced metric on the rod's worldsheet} 
  The induced metric $\bbh=X^* \bbg$ on $\Sigma:=X(\mR\times I)$   has the components 
  $$
  \bbh_{AB}=\bbg_{\mu\nu}(X) \partial_A X^\mu \;\! \partial_BX^\nu \, , \quad A,B\in \{\lambda,\tau\} \, ,
  $$
   given below (blank entries in matrices are $0$): 
  $$
  \bbh_{\tau\tau}=
  {\scaleobj{0.81}{\left[\begin{array}{cccc} 1 \;&\; \partial_\tau \xi \;&\; \partial_\tau \eta \;&\; \partial_\tau\zeta \end{array}\right] }}
   {\scaleobj{0.81}{\left[\begin{array}{cccc} -1 & & & \\ & 1\;&\;\psi(t-z)& \\ &\psi(t-z)\;&\;1 & \\ &&&1\end{array}\right]\left[\begin{array}{c} 1 \\[0.06cm] \partial_\tau\xi \\[0.06cm]\partial_\tau\eta \\[0.06cm] \partial_\tau\zeta \end{array}\right]}}
 \approx -1 \, ,
  $$
  and similarly 
  $ \bbh_{\tau \lambda}=\bbh_{\lambda\tau} \approx  \partial_\tau\xi$ and 
  $ \bbh_{\lambda \lambda} \approx  1+2\partial_\lambda \xi$. Therefore, 
  $$
  [\bbh_{AB}] \approx  {\scaleobj{0.81}{\left[\begin{array}{cc} -1 \;&\; \partial_\tau\xi\\[0.06cm]\partial_\tau\xi\;&\;1+2\partial_\lambda \xi \end{array}\right]}} \, ,
  $$
  with determinant
  $$
  h:=\det [\bbh_{AB}] \approx  -1-2\partial_\lambda \xi \, .
  $$ 
  Also, we have 
  $$
  [\bbh^{AB}]\!:=\![\bbh_{AB}]^{-1}\! \approx  \!
   {\scaleobj{0.81}{\left[\!\begin{array}{cc} -1 \;&\; \partial_\tau\xi\\[0.06cm]\partial_\tau\xi\;&1\!+\!2\partial_\lambda \xi \end{array}\right]^{-1}}}
  \!\!\!=\!{\scaleobj{0.72}{\qfrac{1}{h}}}  
  {\scaleobj{0.81}{\left[\begin{array}{cc} 1\!+\!2\partial_\lambda \xi  \;&\;- \partial_\tau\xi\\[0.06cm]-\partial_\tau\xi\;&\;-1\end{array}\!\right]}}
  \! \approx  \!
   {\scaleobj{0.81}{\left[\!\begin{array}{cc} -1 \;&\; \partial_\tau\xi\\[0.06cm] \partial_\tau\xi \;& 1\!-\!2\partial_\lambda \xi \end{array}\!\right] \, .}}
  $$
  The number density is 
  $$
  n^2:={\scaleobj{0.72}{\qfrac{\bbh_{\tau\tau}}{h}}} \approx  
  {\scaleobj{0.72}{\qfrac{-1}{-1-2\partial_\lambda \xi }}}
   \approx  1-2 \partial_\lambda \xi =1+\delta \, ,
  $$ 
  where 
  $$
  \delta:=-2 \partial_\lambda \xi \, . 
  $$
  As before, given a choice of an elastic law for a non-prestressed rod with longitudinal speed of sound $c\geq 0$, we obtain
  $$
  \rho  \approx  \rho_0 \big( 1 + {\scaleobj{0.9}{\qfrac{\delta}{2}}}\big)
  \;\;\text{ and }\;\; 
  p  \approx  \rho_0 c^2 {\scaleobj{0.9}{\qfrac{\delta}{2}}} \, .
  $$
  The energy momentum tensor $\bbT$ has the components $\bbT^{AB}$ given by 
  \begin{eqnarray*}
  [\bbT^{AB}]\!=\!
  (\rho\!+\!p)\;\![\bbU^A \bbU^B]\!+\!p\;\![\bbh^{AB}]
 \!\!\!\!& \approx &\!\!\!\! 
  (\rho\!+\!p)  
  {\scaleobj{0.81}{\left[\!\begin{array}{cc} 1 & 0\\[0.06cm] 0 & 0\end{array}\!\right]}}
 \! +\!
  p
   {\scaleobj{0.81}{\left[\!\begin{array}{cc} -1 & \partial_\tau\xi\\[0.06cm] \partial_\tau\xi & 1\!-\!2\partial_\lambda \xi  \end{array}\!\right]}}
  \\[0.1cm]
  \!\!\!\!& \approx  &\!\!\!
 \rho_0  {\scaleobj{0.81}{\Bigg[\begin{array}{cc}1+{\scaleobj{0.9}{\qfrac{\delta}{2}}} \;&\; 0 \\ 0 \;&\; {\scaleobj{0.9}{\qfrac{c^2\delta}{2}}}\end{array}\Bigg]}} \, .
  \end{eqnarray*}
 The $x$-component of the equations of motion \eqref{EoM} is then given by
 $$
 \begin{array}{rl}
  0=&\!\!\! {\scaleobj{0.81}{\qfrac{1}{\sqrt{-h}}}} \partial_\tau 
  \pmb{(}\sqrt{-h} \;\!\bbT^{\tau \tau} \partial_\tau x\pmb{)}
  +
  {\scaleobj{0.81}{\qfrac{1}{\sqrt{-h}}}} \partial_\lambda \pmb{(}\sqrt{-h} \;\!\bbT^{\lambda \lambda} \partial_\lambda x\pmb{)}\\[0.27cm]
  &+\;\!2\;\! \bbT^{\tau \tau}  \Gamma^{x}_{ty} ( \partial_\tau t )(  \partial_\tau y) 
  +2\;\! \bbT^{\tau \tau}  \Gamma^{x}_{yz} ( \partial_\tau y )(  \partial_\tau z)\\[0.21cm]
  &+\;\!2\;\! \bbT^{\lambda \lambda}  \Gamma^{x}_{ty} ( \partial_\lambda t )(  \partial_\lambda y) 
  +2\;\! \bbT^{\lambda \lambda}  \Gamma^{x}_{yz} ( \partial_\lambda y )(  \partial_\lambda z) \, ,
 \end{array}
 $$
 that is,
 $$
  \begin{array}{rl}
  0 \approx  &\!\!\!
  {\scaleobj{0.81}{\qfrac{1}{\sqrt{1-\delta}} }}
  \partial_\tau \pmb{\big(}\sqrt{1-\delta} (1+{\scaleobj{0.81}{\qfrac{\delta}{2}}}) \partial_\tau x\pmb{\big)}
  +
  {\scaleobj{0.81}{\qfrac{1}{\sqrt{1-\delta}}}} \partial_\lambda \pmb{\big(}\sqrt{1-\delta} {\scaleobj{0.81}{\qfrac{c^2\delta}{2}}}  \partial_\lambda x\pmb{\big)}\\[0.3cm]
  &+\;\! \big(1+{\scaleobj{0.81}{\qfrac{\delta}{2}}}\big) \psi'(t-z) ( 1 )(  \partial_\tau y) 
  - \big(1+{\scaleobj{0.81}{\qfrac{\delta}{2}}}\big)  
  \psi'(t-z) ( \partial_\tau y )(  \partial_\tau z)\\[0.21cm]
 & +\;\!  {\scaleobj{0.81}{\qfrac{c^2\delta}{2}}} \psi'(t-z) ( 0)(  \partial_\lambda y) 
  -  {\scaleobj{0.81}{\qfrac{c^2\delta}{2}}}  \psi'(t-z)
  ( \partial_\lambda y )(  \partial_\lambda z) \, .
 \end{array}
 $$
 Thus we obtain
 $$
 \partial_\tau^2 \xi +{\scaleobj{0.81}{\qfrac{c^2}{2}}}(\partial_\lambda (-2 \partial_\lambda \xi ))(1+\partial_\lambda \xi) \approx  0 \, ,
 $$ 
 which yields the usual wave equation 
 $$
 \partial_\tau^2 \xi -c^2 \partial_\lambda^2 \xi \approx  0 \, .
 $$
In a similar manner, one can also derive that 
$$
\partial_\tau^2 \eta\;\! \approx  \;\!0\;\;\textrm{ and } \;\;\partial_\tau^2 \zeta\;\! \approx  \;\!0 \, ,
$$
which again trivially correspond to inertial motion of the whole rod along the $y$ or the $z$-axis.

\subsection{The boundary value problem for $\xi$ and for $\partial_\lambda \xi$} 

The boundary conditions are obtained by setting $p=0$ (that is, $\delta=0$) at the endpoints when $\lambda=0$ or $\lambda=L$. This results in the conditions 
$$
\left. \begin{array}{rcl}
(\partial_\lambda \xi)(\tau, 0)\!\!\!&=&\!\!\!\!0\\[0.15cm]
(\partial_\lambda \xi)(\tau, L)\!\!\!&=&\!\!\!\!0 
\end{array}\right\} \textrm{ for all }\tau\in \mR \, .
$$
It is clear that if we assume that $\mR\times [0,L]\mapsto \xi(\tau,\lambda)$ is a solution to the wave equation $\partial_\tau^2 \xi -c^2 \partial_\lambda^2 \xi=0$ with the above  boundary conditions, then $\partial_\lambda\xi$ satisfies the following boundary value problem:
$$
\begin{array}{lll}
\textrm{\bf (PDE)} &  
\partial_\tau^2 (\partial_\lambda \xi) -c^2 \partial_\lambda^2 
(\partial_\lambda \xi)=0 & (\tau \in \mR, \;\lambda \in [0,L] ) \, ,
\\[0.27cm]
\textrm{\bf (BC)} & 
\left\{ \begin{array}{rcl}
(\partial_\lambda \xi)(\tau, 0)\!\!\!&=&\!\!\!\!0\\[0.15cm]
(\partial_\lambda \xi)(\tau, L)\!\!\!&=&\!\!\!\!0
\end{array}\right\} & (\tau\in \mR) \, .
\end{array}
$$
Assuming that the rod is initially at rest (i.e., that the motion of the rod occurs purely in response to the gravitational wave perturbation), we conclude that 
$$
(\partial_\lambda \xi)(\tau,\lambda)
=0
 $$
 for all $\tau \in \mR, \;\lambda \in [0,L]$.
 
 \begin{remark}
What can actually be measured along the rod is the stretching or the pressure, both of which are proportional to
$$
\delta = -2 \partial_\lambda \xi = 0 \, .
$$
In other words, a rod oriented along the $x$-axis does not respond to a cross-polarised gravitational wave, but only to a plus-polarised gravitational wave.
\end{remark}

 \section{Spinning ring encountering a gravitational wave}
 \label{section_ring}

\subsection{Gravitational wave in cylindrical coordinates} 

We now consider the effect of a gravitational wave on a spinning ring. Because of the ring's symmetry under rotations, the effect of a plus-polarised wave and that of a cross-polarised wave will be the same up to a $45^\circ$ rotation, and so we can take the metric $\bbg$ to be that of a plus-polarised wave without loss of generality. In cylindrical coordinates $(t,r,\theta,z)$, $\bbg$ is given by 
$$
\begin{array}{rcl}
ds^2\!\!\!&=&\!\!\!-dt^2 +(1+\varphi(t-z)\cos(2\theta))dr^2 +r^2(1-\varphi(t-z) \cos(2\theta))d\theta^2 
\\[0.06cm]
\!\!\!&&\!\!\!-2\varphi(t-z) \sin (2\theta) rdrd\theta +dz^2
\end{array}
$$
for a wave propagating along the $z$-axis. Thus the matrix of components of $\bbg$ is given as follows (with blank entries being $0$):
$$
[\bbg_{\mu\nu}] 
=
{\scaleobj{0.81}{\left[\begin{array}{cccc}
-1 \;&\;&\;&\;\\
\;&\; 1+\varphi(t-z)\cos(2\theta) \;&\; -\varphi(t-z) \sin(2\theta) r \;&\; \\[0.06cm]
\;&\; -\varphi(t-z) \sin(2\theta) r \;&\; r^2(1-\varphi(t-z)\cos(2\theta)) \;&\; \\
&&\;&\;1
\end{array}\right]}} \, ,
$$
and has the inverse 
$$
[\bbg^{\mu \nu}]
=
{\scaleobj{0.81}{\left[\begin{array}{cccc}
-1 \;&\;&\;&\;\\
\;&\; {\scaleobj{0.9}{\qfrac{1-\varphi(t-z)\cos(2\theta)}{1-(\varphi(t-z))^2}}}\;&\; {\scaleobj{0.9}{\qfrac{\varphi(t-z)\sin(2\theta)}{r(1-(\varphi(t-z))^2)}}} \;&\; \\[0.3cm]
\;&\; {\scaleobj{0.9}{\qfrac{\varphi(t-z)\sin(2\theta)}{r(1-(\varphi(t-z))^2)}}} \;&\; 
{\scaleobj{0.9}{\qfrac{1+\varphi(t-z)\cos(2\theta)}{r^2(1-(\varphi(t-z))^2)}}} \;&\; \\
\;&\;&\;&\;1
\end{array}\right]}} \, .
$$
The nonzero connection coefficients for the Levi-Civita connection $\nabla_\bbg$ induced by $\bbg$ are given (up to first order) as follows:
$$
{\scaleobj{0.96}{
\begin{array}{rcl}
\Gamma^t_{rr}\!\!\!& \approx &\!\!\! {\scaleobj{0.72}{\qfrac{\varphi'(t-z)\cos (2\theta)}{2}}} \, , \\[0.1cm]
\Gamma^t_{r\theta}=\Gamma^t_{\theta r}\!\!\!& \approx &
\!\!\! -{\scaleobj{0.72}{\qfrac{r\varphi'(t-z)\sin (2\theta)}{2}}} \, , \\[0.06cm]
\Gamma^t_{\theta \theta}\!\!\!& \approx &\!\!\! 
-{\scaleobj{0.72}{\qfrac{r^2\varphi'(t-z)\cos (2\theta)}{2}}} \, , \\[0.06cm]
\Gamma^r_{tr}=\Gamma^r_{ rt}\!\!\!& \approx &\!\!\! 
{\scaleobj{0.72}{\qfrac{\varphi'(t-z)\cos (2\theta)}{2}}} \, , \\[0.06cm]
\Gamma^r_{t\theta}=\Gamma^r_{ \theta t}\!\!\!& \approx &\!\!\! 
-{\scaleobj{0.72}{\qfrac{r\varphi'(t-z)\sin (2\theta)}{2}}} \, , \\[0.06cm]
\Gamma^r_{rz}=\Gamma^r_{ zr}\!\!\!& \approx &\!\!\! 
-{\scaleobj{0.72}{\qfrac{\varphi'(t-z)\cos (2\theta)}{2}}} \, , \\[0.06cm]
\Gamma^r_{\theta \theta} \!\!\!& \approx &\!\!\! -r \, , \\[0.06cm]
\Gamma^r_{\theta z}=\Gamma^r_{ z\theta }\!\!\!& \approx &\!\!\! 
{\scaleobj{0.72}{\qfrac{r\varphi'(t-z)\sin (2\theta)}{2}}} \, , \\[0.06cm]
\Gamma^\theta_{tr} =\Gamma^\theta_{rt} \!\!\!& \approx &\!\!\! 
-{\scaleobj{0.72}{\qfrac{\varphi'(t-z)\sin (2\theta)}{2r}}} \, , \\[0.06cm]
\Gamma^\theta_{t \theta} =\Gamma^\theta_{ \theta t} \!\!\!& \approx &\!\!\! 
-{\scaleobj{0.72}{\qfrac{\varphi'(t-z)\cos (2\theta)}{2}}} \, , \\[0.06cm]
\Gamma^\theta_{r \theta} =\Gamma^\theta_{\theta r} \!\!\!& \approx &\!\!\! 
{\scaleobj{0.72}{\qfrac{1}{r}}} \, , \\[0.1cm]
\Gamma^\theta_{rz} =\Gamma^\theta_{zr} \!\!\!& \approx &\!\!\! 
{\scaleobj{0.72}{\qfrac{\varphi'(t-z)\sin (2\theta)}{2r}}} \, , \\[0.06cm]
\Gamma^\theta_{\theta z} =\Gamma^\theta_{z\theta } \!\!\!& \approx &\!\!\!
 {\scaleobj{0.72}{\qfrac{\varphi'(t-z)\cos (2\theta)}{2}}} \, , \\[0.06cm]
\Gamma^z_{rr} \!\!\!& \approx &\!\!\! 
{\scaleobj{0.72}{\qfrac{\varphi'(t-z)\cos (2\theta)}{2}}} \, , \\[0.06cm]
\Gamma^z_{r\theta} =\Gamma^z_{\theta r} \!\!\!& \approx &\!\!\!
 -{\scaleobj{0.72}{\qfrac{r\varphi'(t-z)\sin(2\theta)}{2}}} \, , \\[0.06cm]
\Gamma^z_{\theta\theta}  \!\!\!& \approx &\!\!\! 
-{\scaleobj{0.72}{\qfrac{r^2\varphi'(t-z)\cos (2\theta)}{2} }} \, .
\end{array}}}
$$

 \subsection{{Induced metric on the ring's worldsheet} } 
 
 As usual, we model a spinning ring in the $4$-dimensional spacetime $(M,\bbg)$ by an embedding $X:\mR\times I\rightarrow M$, where $I:=[0,2\pi/k]\subset \mR$ is the interval labelling the points along the ring, $k>0$ is a constant, and $X(\cdot,0)=X(\cdot,2\pi/k)$. As explained in Section~\ref{section_preliminaries}, the parameter $\lambda \in I$ is the arclength in the ring's relaxed configuration in Minkowski spacetime, so that $1/k$ is the radius of the ring in that configuration. If $R>0$ is the radius of the ring when spinning with angular speed $\Omega$, and we assume that the ring is initially lying on the $xy$-plane, then
 $$
 X(\tau,\lambda)=
{\scaleobj{0.81}{\left[\begin{array}{cccc} t(\tau,\lambda)\\[0.06cm]
 r(\tau,\lambda)\\[0.06cm] \theta(\tau,\lambda)\\[0.06cm]
 z(\tau,\lambda)\end{array}\right]}}
 =
 {\scaleobj{0.81}{\left[\begin{array}{cccc} \tau \\[0.06cm] R+\rho(\tau,\lambda)\\[0.06cm] \Omega \tau +k\lambda+\alpha(\tau,\lambda)\\[0.06cm] \zeta(\tau,\lambda)
  \end{array}\right] \, ,}}
  $$
  where $(\tau,\lambda)\mapsto \rho(\tau,\lambda),\;\alpha(\tau,\lambda), \;\zeta(\tau,\lambda)$ describe the small perturbations of the coordinates of the particles along the ring.   The metric $\bbh=X^* \bbg$ on $\Sigma:=X(\mR\times I)$   has the components 
  $$
  \bbh_{AB}=\bbg_{\mu\nu}(X) \partial_A X^\mu \;\! \partial_BX^\nu, \quad A,B\in \{\lambda,\tau\},
  $$
   given below (with blank entries in the $4\times 4$-matrix being $0$):
  $$
 \!\!\!\!\! \begin{array}{rcl}
  \!\!\!\!&&\!\!\!\!\bbh_{\tau\tau}\\[-0.45cm]
  \!\!\!\!\!\!&=&\!\!\!\!\!
 {\scaleobj{0.78}{\left[\!\!\begin{array}{cccc}1 & \partial_\tau \rho & \Omega\!+\!\partial_\tau \alpha & \partial_\tau\zeta \end{array}\!\!\right] \!
  \left[\!\!\begin{array}{cccc}
-1\! &\!\!&&\\
\!&\!\! 1\!+\!\varphi(t\!-\!z)\cos(2\theta) & -\varphi(t\!-\!z) \sin(2\theta) r & \\[0.06cm]
\!&\!\! -\varphi(t\!-\!z) \sin(2\theta) r & r^2(1\!-\!\varphi(t\!-\!z)\cos(2\theta)) & \\
\!&\!\!&&\!\!1
\end{array}\!\!\right]\!
  \left[\!\!\begin{array}{cccc} 1 \\[0.06cm] \partial_\tau \rho \\[0.06cm] \Omega\!+\!\partial_\tau \alpha \\[0.06cm] \partial_\tau\zeta \end{array}\!\!\right]}} 
  \\[0.48cm]
  \!\!\!\!&  \approx  &\!\!\! -1+R^2\Omega^2+R \Omega(2\partial_\tau \theta+2\Omega \rho -\varphi(t-z)\cos(2(\Omega \tau +k\lambda))) \, ,
  \end{array}
  $$
  and similarly 
  $$
  \begin{array}{rcl}
  \bbh_{\tau \lambda}\!=\!\bbh_{\lambda\tau}
   \!\!\!\!& \approx &\!\!\!\!
  kR^2 \Omega\!+\!R^2(k \partial_\tau \alpha\! +\!\Omega \partial_\lambda \alpha\!-\!k\Omega \varphi \cos(2(\Omega \tau \!+\!k\lambda)) \!+\!2R\Omega k \rho,\\[0.1cm]
   \bbh_{\lambda \lambda}\!\!\!\!& \approx &\!\!\! \!
  k^2 R^2 \!+\!kR^2 (2\partial_\lambda \alpha\! -\!k\varphi(t\!-\!z) \cos(2(\Omega \tau +k\lambda)))\!+\!2Rk^2 \rho \, .
  \end{array}
  $$
  So 
  $
  [\bbh_{AB}] \approx 
 {\scaleobj{0.78}{ \left[\begin{array}{cc} 
  -1+R^2\Omega^2+A
 \; &\; 
  kR^2 \Omega+B \\[0.06cm]
  kR^2 \Omega+B
  \;&\; 
  k^2 R^2 +C
  \end{array}\right]}} \, ,
  $
  where 
  \begin{eqnarray*}
  A\!\!\!&:=&\!\!\!R \Omega(2\partial_\tau \theta+2\Omega \rho -\varphi (t-z)\cos(2(\Omega \tau +k\lambda))) \, ,\\
  B\!\!\!&:=&\!\!\!
  R^2(k \partial_\tau \alpha +\Omega \partial_\lambda \alpha-k\Omega \varphi (t-z)\cos(2(\Omega \tau +k\lambda)) +2R\Omega k \rho \, ,\\
  C\!\!\!&:=&\!\!\!
  kR^2 (2\partial_\lambda \alpha -k\varphi(t-z) \cos(2(\Omega \tau +k\lambda)))+2Rk^2 \rho \, .
  \end{eqnarray*}
   Hence 
  $$
  h\!=\!\det [\bbh_{AB}]
  \! \approx \! 
  -k^2 R^2\!+\!k^2 R^2 A\!-\!2kR^2\Omega B\!+\!(R^2\Omega^2\!-\!1)C
  \!=\! -k^2 R^2 (1\!-\!D) \, ,
  $$
  where 
   $
  D:= A-{\scaleobj{0.72}{\qfrac{2\Omega }{k}}}B+
  {\scaleobj{0.72}{\qfrac{R^2 \Omega^2-1}{k^2 R^2}}} C.
  $ 
  Also, we have 
 $$
 \begin{array}{rcl}
  [\bbh^{AB}]\!:=\![\bbh_{AB}]^{-1}
  \!\!\!\!\!\!& \approx &\!\!\!\!\!
 -{\scaleobj{0.72}{\qfrac{1}{k^2 R^2}}}
  {\scaleobj{0.78}{ \left[\!\begin{array}{cc} 
  k^2 R^2 \!+\!C\!+\!k^2 R^2 D &
  -kR^2 \Omega \!-\!B\!-\!kR^2 \Omega D \\[0.06cm]
  -kR^2 \Omega \!-\!B\!-\!kR^2 \Omega D &
  -1\!+\!R^2 \Omega^2 \!+\!A \!+\!(-1\!+\!R^2 \Omega^2)D
  \end{array}\!\right]}}
  \\[0.3cm]
  \!\!\!&=&\!\!\! {\scaleobj{0.78}{ \Bigg[\begin{array}{cc}
    -1 + \delta_{11}  \;&\; \frac{\Omega}{k}+\delta   \\[0.06cm]
  {\scaleobj{0.9}{\qfrac{\Omega}{k}}}+\delta\;&\; n_0^2 (1+\Delta)
  \end{array}\Bigg] \, , }}
  \end{array} 
 $$
 where 
 $$
 \begin{array}{rcl}
 \delta_{11}\!\!\!&:=&\!\!\! - {\scaleobj{0.9}{\qfrac{1}{k^2 R^2}}} C-D \, , \\[0.15cm]
 \delta \!\!\!&:=&\!\!\!  {\scaleobj{0.9}{\qfrac{1}{k^2 R^2 }}} B +{\scaleobj{0.72}{\qfrac{\Omega}{k}}} D \, , \\[0.15cm]
 n_0^2  \!\!\!&:=&\!\!\! {\scaleobj{0.9}{\qfrac{1-R^2 \Omega^2}{k^2 R^2}}} \, , \\[0.15cm]
 \Delta \!\!\!&:=&\!\!\! -{\scaleobj{0.9}{\qfrac{1}{1-R^2 \Omega^2}}} A+D \, .
 \end{array}
 $$
 \begin{spacing}{1.3} 
 \noindent The number density is  
  $
 n^2 ={\scaleobj{0.72}{\qfrac{\bbh_{\tau \tau}}{h}}}=\bbh^{\tau \tau} \approx   n_0^2 (1+\Delta).
 $ 
 As the ring is rotating, the unperturbed state (not to be confused with the relaxed state) does not have zero pressure in general. If 
 \end{spacing}
 
 \vspace{-0.6cm}
 
 $$
 \rho_0=F(n_0^2), \quad p_0=2n_0^2F'(n_0^2)-F(n_0^2), \quad c^2 = 2n_0^2{\scaleobj{0.9}{\qfrac{F''(n_0^2)}{F'(n_0^2)}}} + 1
 $$ 
 designate the density, the pressure and the longitudinal speed of sound in the unperturbed state, then one can easily see that
 $$
 \rho \;\! \approx \;\! \rho_0 + {\scaleobj{0.9}{\qfrac{\rho_0+p_0}{2n_0^2}}} (n^2 - n_0^2)  \approx   \rho_0 + {\scaleobj{0.9}{\qfrac{\rho_0+p_0}{2}}} \Delta 
 $$
 and
  $$
 p \;\! \approx \;\! p_0 + {\scaleobj{0.9}{\qfrac{(\rho_0+p_0)c^2}{2n_0^2}}} (n^2 - n_0^2)  \approx  p_0 + {\scaleobj{0.9}{\qfrac{(\rho_0+p_0)c^2}{2}}} \Delta \, .
 $$
 The  components of the energy momentum tensor $\bbT$ are then given by 
  $$
  \begin{array}{rcl}
  \bbT^{\tau\tau}
  \!\!\!&  \approx &\!\!\!
  \rho_0 + {\scaleobj{0.9}{\qfrac{\rho_0+p_0}{2}}} \Delta + p_0 \delta_{11} \, ,\\
  \bbT^{\tau\lambda}=\bbT^{\lambda \tau}
  \!\!\!&  \approx &\!\!\!
 p_0{\scaleobj{0.9}{\qfrac{\Omega}{k}}} 
 + {\scaleobj{0.9}{\qfrac{(\rho_0+p_0)c^2}{2} \qfrac{\Omega}{k}}} \Delta + p_0 \delta \, ,\\
  \bbT^{\lambda \lambda}
  \!\!\!&  \approx &\!\!\!
  p_0 n_0^2 + {\scaleobj{0.9}{\qfrac{n_0^2}{2}}} \left( \rho_0 c^2 + (2+c^2) p_0 \right) \Delta \, .
  \end{array}
  $$ 
    Using the expression for the transverse speed of sound in the unperturbed state,
  $$
  s = \sqrt{{\scaleobj{0.9}{-\qfrac{p}{\rho}}}} \, ,
  $$
  the  energy momentum tensor $\bbT$ components can also be written as 
  $$
  \begin{array}{rcl}
  \bbT^{\tau\tau}
  \!\!\!&  \approx &\!\!\!
  \rho_0 \pmb{\big(} 1 + {\scaleobj{0.9}{\qfrac{1-s^2}{2}}} \Delta - s^2 \delta_{11} \pmb{\big)} \, ,\\
  \bbT^{\tau\lambda}=\bbT^{\lambda \tau}
  \!\!\!&  \approx &\!\!\!
 \rho_0 \pmb{\big(} -s^2 {\scaleobj{0.9}{\qfrac{\Omega}{k}}}
  + {\scaleobj{0.9}{\qfrac{(1-s^2)c^2}{2} \qfrac{\Omega}{k} }}
  \Delta - s^2 \delta \pmb{\big)} \, ,\\
  \bbT^{\lambda \lambda}
  \!\!\!&  \approx &\!\!\!
  \rho_0 \pmb{\big(} -s^2 n_0^2 + {\scaleobj{0.9}{\qfrac{n_0^2}{2}}} \big( c^2 - (2+c^2) s^2 \big) \Delta \pmb{\big)} \, .
  \end{array}
  $$ 
 
\subsection{Linearised equations of motion}

 The equations of motion can now be obtained in exactly the same manner as in sections~\ref{section_rod_plus} and \ref{section_rod_cross}, but the calculations are much more cumbersome and will not be shown here. To zeroth order, they yield the equilibrium condition
  $$
  s^2 = \Omega^2 R^2
  $$  
  (which is simply the equilibrium condition in Minkowski's spacetime, see \cite{NQV}), and to first order we obtain the following linearised equations:
 \begin{eqnarray}
  \nonumber
  0	\!\!\!\!\!& \approx  &\!\!\!\!\! 
  {\scaleobj{0.96}{
 -{\scaleobj{0.75}{\qfrac{c^2 (1\!-\!s^2) \Omega}{k^2}}}\partial_\lambda^2\alpha
 -{\scaleobj{0.75}{\qfrac{2 c^2 s^2}{k}}}\partial_\lambda \partial_\tau\alpha
 -{\scaleobj{0.75}{\qfrac{\Omega^2 (c^2\!+\!s^2)}{k s}}}\partial_\lambda\rho
 +{\scaleobj{0.75}{\qfrac{s^2(1\!-\!c^2 s^2)}{\Omega(1\!-\!s^2)}}}\partial_\tau^2\alpha }}
 \\[0.06cm] 
 \label{lin_eq_3}	&&
	\!\!\!\!
	{\scaleobj{0.96}{
	+{\scaleobj{0.75}{\qfrac{s \Omega  (2\!-\!c^2\!-\!s^2)}{1\!-\!s^2}}} \partial_\tau\rho\!
	+\!{\scaleobj{0.75}{\qfrac{s^2 (c^2\!+\!s^2\!-\!2)\cos (2 (\Omega \tau\!+\!k \lambda))}{2(1-s^2)}}}\varphi'\!+\!
	{\scaleobj{0.75}{\qfrac{\Omega  (s^2\!-\!c^2) \sin (2 (\Omega \tau\!+\!k \lambda))}{1\!-\!s^2} }}\varphi}} \, ,
	\phantom{aaa} \\[0.06cm]
	 \nonumber
0	\!\!\!\!& \approx  &\!\!\!\!
	{\scaleobj{0.96}{
{\scaleobj{0.75}{\qfrac{\Omega  (c^2\!+\!s^2)}{k s}}}\partial_\lambda\alpha
\!-\!{\scaleobj{0.75}{\qfrac{s (2\!-\!c^2\!-\!s^2)}{1\!-\!s^2}}}\partial_\tau\alpha
\!-\!{\scaleobj{0.75}{\qfrac{\Omega^2 (s^2\!-\!c^2)}{(1\!-\!s^2) s^2}}}\rho
\!-\!{\scaleobj{0.75}{\qfrac{(1\!-\!s^2) \Omega^2}{k^2}}}\partial_\lambda^2\rho
\!-\!{\scaleobj{0.75}{\qfrac{2 s^2 \Omega}{k}}}\partial_\lambda \partial_\tau\rho}}\\[0.06cm] 
	 \label{lin_eq_4}&&\!\!\!\!
	{\scaleobj{0.96}{
+(s^2\!+\!1) \partial_\tau^2\rho\!-\!\tfrac{\Omega (c^2\!-\!s^2) \cos (2 (\Omega \tau\!+\!k \lambda))}{2 (1-s^2) s}\varphi
-s \sin (2 (\Omega \tau\!+\!k \lambda )) \varphi' }} \, , \phantom{aaa}\\[0.06cm] 
	 \nonumber
 0	\!\!\!\!& \approx  &\!\!\!\! 
	{\scaleobj{0.96}{-
{\scaleobj{0.75}{\qfrac{c^2 (1\!-\!s^2) \Omega^2}{k^2 s^2}}}\partial_\lambda^2\alpha
\!-\!{\scaleobj{0.75}{\qfrac{2 c^2 \Omega  }{k}}}\partial_\lambda \partial_\tau\alpha
\!-\!{\scaleobj{0.75}{\qfrac{\Omega^3 (c^2\!+\!s^2)}{k s^3}}}\partial_\lambda \rho
\!+\!{\scaleobj{0.75}{\qfrac{(1\!-\!c^2 s^2)}{1\!-\!s^2}}}\partial_\tau^2\alpha}}\\[0.06cm] 
	 \label{lin_eq_5}&&\!\!\!\!
	{\scaleobj{0.96}{
+{\scaleobj{0.75}{\qfrac{\Omega^2 (2\!-\!c^2\!-\!s^2) }{(1\!-\!s^2)s}}}
\partial_\tau\rho\!
-\!{\scaleobj{0.75}{\qfrac{\Omega  (2\!-\!c^2\!-\!s^2) \cos (2 (\Omega \tau\!+\!k \lambda))}{2 (1\!-\!s^2)}}}\varphi'\!
+\!{\scaleobj{0.75}{\qfrac{\Omega^2 (s^2\!-\!c^2) \sin (2 (\Omega \tau\!+\!k \lambda))}{(1\!-\!s^2) s^2}}} \varphi}} \, ,
	\phantom{aaa}\\[0.06cm] 
	 \label{lin_eq_6}
	  0	\!\!\!\!& \approx  &\!\!\!\! 
	  	{\scaleobj{0.96}{
{\scaleobj{0.75}{\qfrac{\Omega^2(1\!-\!s^2)}{k^2}}}\partial^2_\lambda\zeta	
\!-\!(1+s^2)\partial^2_\tau \zeta
\!+\!{\scaleobj{0.75}{\qfrac{2 \Omega s^2  }{k}}} \partial_\lambda \partial_\tau\zeta}} \, .
 \phantom{aaa}
 \end{eqnarray}
 Notice that the first and the third equations are the same. This is to be expected, as the equation along $\frac{\partial\hfill}{\partial \tau}$ is automatically satisfied (see \cite{NQV}).
 
 \subsection{Non-rotating case}
 
 Let us consider first the case of a non-rotating ring, corresponding to $\Omega=s=0$. In this limit we have, from the equilibrium condition, $s=k\Omega$, and so the equations of motion become
$$
\begin{array}{rcl}
 -c^2\partial_\lambda^2\alpha-c^2k^2\partial_\lambda\rho+\partial_\tau^2\alpha-c^2 k^2 \sin \left(2 k \lambda\right) \varphi \!\!\!\!& \approx  &\!\!\!\! 0 \, ,\\[0.18cm]
c^2\partial_\lambda\alpha+c^2k^2\rho+\partial_\tau^2\rho-\tfrac12 c^2 k \cos \left(2 k \lambda\right)\varphi \!\!\!\!& \approx  &\!\!\!\! 0 \, ,\\[0.18cm]
 \partial^2_\tau \zeta \!\!\!\!& \approx  &\!\!\!\!  0 \, .
 \end{array}
 $$
 The last equation is trivial, and corresponds to inertial motion of the whole ring along the $z$-axis. The other two equations are coupled, and can be solved by decomposing $\rho$ and $\alpha$ in Fourier series:
 $$
 \textstyle 
 \rho(\tau,\lambda) = \sum\limits_{m\in\mathbb{Z}} c_m(\tau) e^{imk\lambda} \, ,
 \qquad
 \alpha(\tau,\lambda) = \sum\limits_{m\in\mathbb{Z}} d_m(\tau) e^{imk\lambda} \, .
 $$
 Substituting into the second equation yields
 $$
  \textstyle 
   \sum\limits_{m\in\mathbb{Z}} \! \pmb{(} im kc^2 d_m(\tau) + c^2 k^2 c_m(\tau) + \ddot{c}_m(\tau) \pmb{)} e^{imk\lambda}\! =\! \tfrac1{4} c^2 k (e^{2i k \lambda} + e^{-2i k \lambda} )\varphi(\tau) \, ,
 $$
 while substituting into the first equation yields
 $$
\textstyle 
 \sum\limits_{m\in\mathbb{Z}} \!\pmb{(} c^2 m^2 k^2 d_m(\tau) - im c^2 k^3 c_m(\tau) + \ddot{d}_m(\tau) \pmb{)} e^{imk\lambda} \!=\! \tfrac{c^2 k^2}{2i} (e^{2i k \lambda} - e^{-2i k \lambda} )\varphi(\tau) \, .
 $$
Thus we get the system of second order ordinary differential equations
$$
\left\{\begin{array}{rcl}
 \ddot{c}_m(\tau) + c^2 k^2 c_m(\tau) + im kc^2 d_m(\tau) \!\!\!\!&=&\!\!\! 0 \\[0.12cm]
 \ddot{d}_m(\tau) + c^2 m^2 k^2 d_m(\tau) - im c^2 k^3 c_m(\tau) \!\!\!\!&=&\!\!\! 0
\end{array}\right.
$$
for $|m|\neq 2$, together with
$$
\left\{\begin{array}{rcl}
 \ddot{c}_2(\tau) + c^2 k^2 c_2(\tau) + 2i kc^2 d_2(\tau) \!\!\!\!&=&\!\!\! \tfrac1{4} c^2 k \varphi(\tau) \\[0.12cm]
 \ddot{d}_2(\tau) + 4 c^2 k^2 d_2(\tau) - 2i c^2 k^3 c_2(\tau) \!\!\!\!&=&\!\!\! -\tfrac{i}{2} c^2 k^2 \varphi(\tau)
\end{array}\right.
$$
and
$$
\left\{\begin{array}{rcl}
 \ddot{c}_{-2}(\tau) + c^2 k^2 c_{-2}(\tau) - 2i kc^2 d_{-2}(\tau) \!\!\!\!&=&\!\!\! \tfrac1{4} c^2 k \varphi(\tau) \\[0.12cm]
 \ddot{d}_{-2}(\tau) + 4 c^2 k^2 d_{-2}(\tau) + 2i c^2 k^3 c_{-2}(\tau) \!\!\!\!&=&\!\!\! \tfrac{i}{2} c^2 k^2 \varphi(\tau)\, .
\end{array}\right.
$$
In other words, only the modes $m=2$ and $m=-2$ are excited by the gravitational wave. If we assume that the motion of the ring occurs purely in response to the gravitational wave perturbation, then we can set $c_m(\tau)=d_m(\tau)=0$ for $|m|\neq 2$. To solve the system of ordinary differential equations for the $m=2$ mode, we take a Fourier transform in time. Writing
$$
\begin{array}{l}
c_2(\tau) \!=\! {\scaleobj{0.9}{\qfrac{1}{2\pi}\displaystyle \int_{-\infty}^{\infty}}} \!\widehat{c_2}(\omega) e^{i\omega \tau} d\omega \, ,
\\[0.33cm]
d_2(\tau) \!=\! {\scaleobj{0.9}{\qfrac{1}{2\pi}\displaystyle \int_{-\infty}^{\infty}}}\!
 \widehat{d_2}(\omega) e^{i\omega \tau} d\omega \, ,
\\[0.33cm]
\varphi(\tau)\! = \!{\scaleobj{0.9}{\qfrac{1}{2\pi}\displaystyle \int_{-\infty}^{\infty} }}\!\widehat{\varphi}(\omega) e^{i\omega \tau} d\omega \, ,
\end{array}
$$
we obtain
$$
\left\{\begin{array}{rcl}
 -\omega^2 \widehat{c_2}(\omega) + c^2 k^2  \widehat{c_2}(\omega) + 2i kc^2  \widehat{d_2}(\omega)  \!\!\!\!&=&\!\!\!  \tfrac1{4} c^2 k \widehat{\varphi}(\omega) \\[0.12cm]
 -\omega^2 \widehat{d_2}(\omega) + 4 c^2 k^2 \widehat{d_2}(\omega) - 2i c^2 k^3 \widehat{c_2}(\omega)  \!\!\!\!&=&\!\!\!  -\tfrac{i}{2} c^2 k^2 \widehat{\varphi}(\omega) \, ,
\end{array}\right.
$$
that is,
$$
{\scaleobj{0.9}{\bigg[
\begin{matrix}
-\omega^2 + c^2 k^2 & 2i kc^2 \\[0.1cm]
- 2i c^2 k^3 & -\omega^2 + 4 c^2 k^2
\end{matrix}
\bigg]
\bigg[
\begin{matrix}
\widehat{c_2}(\omega) \\[0.1cm]
\widehat{d_2}(\omega) 
\end{matrix}
\bigg]}}
=
{\scaleobj{0.9}{\bigg[
\begin{array}{r}
\tfrac1{4} c^2 k  \\[0.1cm]
-\tfrac{i}{2} c^2 k^2 
\end{array}
\bigg] \widehat{\varphi}(\omega) \, .}}
$$
The determinant of the system's matrix, $A(\omega)$, is
$$
\det A(\omega) = \omega^2 (\omega^2 - 5c^2k^2) \, ,
$$
and so for $\omega^2 \neq 0, 5c^2k^2$, we have necessarily
$$
{\scaleobj{0.9}{\bigg[
\begin{array}{r}
\widehat{c_2}(\omega) \\[0.1cm]
\widehat{d_2}(\omega) 
\end{array}
\bigg]}}
=
{\scaleobj{0.9}{\bigg[
\begin{array}{r}
-\tfrac1{4} c^2 k  \\[0.1cm]
\tfrac{i}{2} c^2 k^2 
\end{array}
\bigg] \qfrac{\widehat{\varphi}(\omega)}{\omega^2 - 5c^2k^2} \, .}}
$$
The resonant frequencies for the non-rotating ring, where the response to the gravitational wave signal will be stronger, are then given by
$$
\omega = \pm \sqrt{5}ck \, .
$$

To the particular solution obtained above, we can add the solutions of the homogeneous system, which arise precisely from Dirac delta functions at the excluded values of $\omega$. For example, if we substitute $\widehat{c_2}(\omega)=C_2 \delta(\omega-\sqrt{5}ck)$ and $\widehat{d_2}(\omega)=D_2 \delta(\omega-\sqrt{5}ck)$ into the homogeneous equation,we obtain
$$ 
{\scaleobj{0.9}{\bigg[
\begin{matrix}
-4c^2 k^2 & 2i kc^2 \\[0.1cm]
- 2i c^2 k^3 & -c^2 k^2
\end{matrix}
\bigg]
\bigg[
\begin{matrix}
C_2 \\[0.1cm]
D_2 
\end{matrix}
\bigg]}}
={\scaleobj{0.9}{0}}
\Leftrightarrow
{\scaleobj{0.9}{D_2 = -2ik C_2 \, ,}}
$$
\begin{spacing}{1.15}
\noindent 
and the same result holds if we instead set $\widehat{c_2}(\omega)=C_2 \delta(\omega+\sqrt{5}ck)$ and $\widehat{d_2}(\omega)=D_2 \delta(\omega+\sqrt{5}ck)$. Therefore, there are particular solutions of the form
\end{spacing}

\vspace{-0.6cm}

$$
{\scaleobj{0.9}{
\bigg[
\begin{matrix}
\widehat{c_2}(\omega) \\[0.1cm]
\widehat{d_2}(\omega) 
\end{matrix}
\bigg]}}
\!=\!
{\scaleobj{0.9}{\bigg[\!\!
\begin{array}{r}
-\tfrac1{4} c^2 k  \\[0.1cm]
\tfrac{i}{2} c^2 k^2 
\end{array}
\!\!\bigg] 
\Big(
\frac{\widehat{\varphi}(\omega)}{\omega^2 \!- 5c^2k^2} + E \delta(\omega-\sqrt{5}ck) + F \delta(\omega+\sqrt{5}ck)
\Big) \, ,}}
$$
with $E,F \in \mathbb{C}$ constants. Assuming that $\widehat{\varphi}(\omega)$ is a continuous function, we can rewrite these particular solutions in the form
$$
{\scaleobj{0.9}{
\bigg[
\begin{matrix}
\widehat{c_2}(\omega) \\[0.1cm]
\widehat{d_2}(\omega) 
\end{matrix}
\bigg]}}
\!=\!
{\scaleobj{0.9}{
\bigg[\!\!
\begin{array}{r}
-\tfrac1{4} c^2 k  \\[0.1cm]
\tfrac{i}{2} c^2 k^2 
\end{array}
\!\!\bigg] 
\Big(
\frac{1}{\omega^2\! - 5c^2k^2} + E \delta(\omega-\sqrt{5}ck) + F \delta(\omega+\sqrt{5}ck)
\Big) \widehat{\varphi}(\omega) \, ,}}
$$
for $E,F \in \mathbb{C}$ (possibly different) constants.

Let ${\rm p.v.} \, \frac1{\omega}$ denote the distribution given by  
$$
\textstyle
\big\langle {\rm p.v.} \, 
{\scaleobj{0.9}{\qfrac{1}{\omega}}},\varphi \big\rangle
:=\lim\limits_{\epsilon\rightarrow 0}
{\scaleobj{0.9}{\displaystyle \int_{|\omega|>\epsilon}\qfrac{\varphi(\omega)}{\omega}}}\;\!d\omega
 \,\,  \textrm{ for all } \,\, \varphi\in \calD(\mR) \, .
$$
Then ${\rm p.v.} \, \frac1{\omega}\in \calS'(\mR)$, and its inverse Fourier transform is given by 
$$
\mathcal{F}^{-1} \big( {\rm p.v.} \, {\scaleobj{0.9}{\qfrac{1}{\omega}}}\big) 
= iH(\tau) -{\scaleobj{0.9}{\qfrac{i}{2}}} \, ,
$$
where $H$ denotes the Heaviside step function.
For $a\in \mR$, define the shift operator $S_a :\calS'(\mR)\rightarrow \calS'(\mR)$  by $\langle S_a T, \varphi\rangle=\langle T, \varphi(\cdot+a)\rangle$, $T\in \calS'(\mR)$, $\varphi\in \calS(\mR)$. Then 
$$
{\scaleobj{0.9}{\qfrac{1}{\omega^2 - 5c^2k^2}}}
:= {\scaleobj{0.9}{\qfrac{1}{2\sqrt{5} ck} }}
( S_{\sqrt{5} c k} - S_{-\sqrt{5} c k} )\;\!{\rm p.v.} \, {\scaleobj{0.9}{\qfrac{1}{\omega}}}\in \calS'(\mR) \, .
$$
Consider the tempered distribution
$$
\widehat{\Phi}(\omega)={\scaleobj{0.9}{\qfrac{1}{\omega^2 - 5c^2k^2}}}
 + E \delta(\omega-\sqrt{5}ck) + F \delta(\omega+\sqrt{5}ck) \, .
$$
 Using the well-known results
$$
\mathcal{F}^{-1}  ( \delta(\omega) ) = {\scaleobj{0.9}{\qfrac{1}{2\pi}}}\,,\;\;\textrm{ and }
\;\;
\mathcal{F}^{-1}  ( \widehat{f}(\omega-a) )= e^{ia\tau} f(\tau) \, ,
$$
we obtain, after a partial fraction decomposition,
$$
\Phi(\tau) = - {\scaleobj{0.9}{\qfrac{\sin(\sqrt{5}ck\tau)}{\sqrt{5}ck}}}
 \big( H(\tau) - {\scaleobj{0.9}{\qfrac{1}{2}}} \big) 
 + {\scaleobj{0.9}{\qfrac{E}{2\pi}}} e^{i\sqrt{5}ck\tau} 
 + {\scaleobj{0.9}{\qfrac{F}{2\pi}}} e^{-i\sqrt{5}ck\tau} \, .
$$
Choosing
 $
E = -F = {\scaleobj{0.9}{\qfrac{i\pi}{2\sqrt{5}ck}}} \, ,
$ 
we obtain
 $$
\Phi(\tau) = - {\scaleobj{0.9}{\qfrac{\sin(\sqrt{5}ck\tau)}{\sqrt{5}ck}}} H(\tau) \, ,
$$ 
corresponding to the tempered distribution
$$
\widehat{\Phi}(\omega)={\scaleobj{0.9}{\qfrac{1}{\omega^2 - 5c^2k^2}}}
 +  {\scaleobj{0.9}{\qfrac{i\pi}{2\sqrt{5}ck}}} \delta(\omega-\sqrt{5}ck) 
 - {\scaleobj{0.9}{\qfrac{i\pi}{2\sqrt{5}ck} }}\delta(\omega+\sqrt{5}ck) \, .
$$
This is the distribution yielding the particular solution that corresponds to a ring initially at rest, since it leads to the solution
$$
{\scaleobj{0.9}{
\bigg[
\begin{matrix}
c_2(\tau) \\[0.12cm]
d_2(\tau) 
\end{matrix}
\bigg]}}
=
{\scaleobj{0.9}{
\bigg[
\begin{matrix}
-\tfrac1{4} c^2 k  \\[0.12cm]
\tfrac{i}{2} c^2 k^2 
\end{matrix}
\bigg] 
(\Phi \ast \varphi)(\tau)}}
={\scaleobj{0.9}{
\bigg[
\begin{matrix}
\tfrac1{4} c^2 k  \\[0.12cm]
-\tfrac{i}{2} c^2 k^2 
\end{matrix}
\bigg] 
\displaystyle \int_0^\infty \qfrac{\sin(\sqrt{5}cks)}{\sqrt{5}ck} \varphi(\tau - s) ds \, ,}}
$$
which vanishes for $\tau$ smaller than the infimum of the support of $\varphi(\tau)$ (assumed to be finite).

The calculation for the mode $m=-2$ is very similar, and results in the particular solution
$$
{\scaleobj{0.9}{
\bigg[
\begin{matrix}
c_{-2}(\tau) \\[0.12cm]
d_{-2}(\tau) 
\end{matrix}
\bigg]}}
=
{\scaleobj{0.9}{
\bigg[
\begin{matrix}
-\tfrac1{4} c^2 k  \\[0.12cm]
-\tfrac{i}{2} c^2 k^2 
\end{matrix}
\bigg] 
(\Phi \ast \varphi)(\tau)}}
=
{\scaleobj{0.9}{
\bigg[
\begin{matrix}
\tfrac1{4} c^2 k  \\[0.12cm]
\tfrac{i}{2} c^2 k^2 
\end{matrix}
\bigg] 
\int_0^\infty \qfrac{\sin(\sqrt{5}cks)}{\sqrt{5}ck} \varphi(\tau - s) ds \, .}}
$$
As one would expect, $c_{-2}(t)=\overline{c_2(t)}$ and $d_{-2}(t)=\overline{d_2(t)}$, where the bar denotes complex conjugation.

\subsection{Rotating case}
Now we consider  a rotating ring, i.e., $\Omega\neq 0$.  Equation \eqref{lin_eq_6} for $\zeta(\tau,\lambda)$ is decoupled from the others, and is insensitive to the presence of the gravitational wave. Assuming that the ring is initially in the equilibrium configuration (i.e., that the perturbations of the ring occurs purely in response to the gravitational wave perturbation), we conclude that 
$$
\zeta(\tau,\lambda)
=0
 $$
 for all $\tau \in \mR, \;\lambda \in [0,2\pi/k]$.
 
The other two equations, namely \eqref{lin_eq_4} and \eqref{lin_eq_3} (which is the same as equation \eqref{lin_eq_5}),  are coupled, and, as before, are analysed by decomposing $\rho$ and $\alpha$ in Fourier series:
 $$
 \begin{array}{rcl}
 \textstyle 
 \rho(\tau,\lambda) \!\!\!&=&\!\!\! \sum\limits_{m\in\mathbb{Z}} c_m(\tau) e^{imk\lambda} \, , \\[0.39cm]
 \alpha(\tau,\lambda) \!\!\!&=&\!\!\!  \sum\limits_{m\in\mathbb{Z}} d_m(\tau) e^{imk\lambda} \, .
 \end{array}
 $$
 Substituting in  \eqref{lin_eq_4}  yields 
 $$
 \!\!\!
 \begin{array}{l}
 {\scaleobj{0.81}{\sum\limits_{m\in \mZ}}}
 {\scaleobj{0.83}{\pmb{(} 
 \tfrac{\Omega(c^2+s^2)}{s} imd_m 
 \!-\! \tfrac{s(2-c^2-s^2)}{1-s^2} \dot{d}_m
 \!-\!\tfrac{\Omega^2 (s^2-c^2)}{(1-s^2)s^2} c_m
 \!+\!(1\!-\!s^2)\Omega^2 m^2 c_m 
\!-\!2s^2 \Omega im \dot{c}_m 
\!+\!(s^2\!+\!1)\ddot{c}_m
 \pmb{)}}}
 \\[0.3cm]
 ={\scaleobj{0.87}{\Omega(c^2-s^2) \tfrac{e^{2\Omega \tau i} e^{2k\lambda i} +e^{-2\Omega \tau i} e^{-2k \lambda i}  }{4(1-s^2)s} \varphi(\tau)
 +s\tfrac{e^{2\Omega \tau i} e^{2k\lambda i} -e^{-2\Omega \tau i} e^{-2k\lambda i} }{2i} \varphi'(\tau)}} \, ,
\end{array}
$$ 
while substituting into \eqref{lin_eq_3} yields
 $$
 \begin{array}{l}
 {\scaleobj{0.87}{\sum\limits_{m\in \mZ}\! \pmb{(}
 \tfrac{c^2 (1-s^2) \Omega^2}{s^2}  m^2 d_m
 \!-\!2c^2\Omega im \dot{d}_m -\tfrac{\Omega^3(c^2+s^2)}{s^3} im c_m
 \!+\!\tfrac{1-c^2 s^2}{1-s^2}\ddot{d}_m
 \!+\!\tfrac{\Omega^2 (2-c^2-s^2)}{(1-s^2)s} \dot{c}_m \;\!\pmb{)}}} \\[0.3cm]
 ={\scaleobj{0.87}{\Omega (2-c^2-s^2) \tfrac{e^{2\Omega \tau i} e^{2k\lambda i}+e^{-2\Omega \tau i}e^{-2k\lambda i}}{4(1-s^2)}\varphi'(\tau)
 -\Omega^2 (s^2-c^2) \tfrac{ e^{2\Omega \tau i} e^{2k\lambda i}-e^{-2\Omega \tau i}e^{-2k\lambda i}}{2i(1-s^2)s^2}\varphi(\tau)\, .}}
\end{array}
$$ 
Thus we obtain the system of second order ordinary differential equations
$$
\left\{\!\!
\begin{array}{rcl}
{\scaleobj{0.83}{0}}\!\!\!\!\!&=&\!\!\!\! {\scaleobj{0.83}{
\frac{\Omega(c^2+s^2)}{s} imd_m 
 \!-\! \frac{s(2-c^2-s^2)}{1-s^2} \dot{d}_m
 \!+\!
 \Omega^2((1\!-\!s^2)m^2\!-\!\frac{s^2-c^2}{(1-s^2)s^2}  ) 
 c_m 
\!-\!2s^2 \Omega im \dot{c}_m 
\!+\!(s^2\!+\!1)\ddot{c}_m }}
\\[0.2cm]
{\scaleobj{0.87}{0}}\!\!\!\!\!&=&\!\!\!\! 
 {\scaleobj{0.87}{\frac{c^2 (1-s^2) \Omega^2}{s^2}  m^2 d_m
 \!-\!2c^2\Omega im \dot{d}_m -\frac{\Omega^3(c^2+s^2)}{s^3} im c_m
 \!+\!\frac{1-c^2 s^2}{1-s^2}\ddot{d}_m
 \!+\!\frac{\Omega^2 (2-c^2-s^2)}{(1-s^2)s} \dot{c}_m}}
\end{array}
\right.
$$
for $|m|\neq 2$, together with 
$$
\left\{\!\!
\begin{array}{rcl}
 \!\!\!\!\!&&\!\!\!\! 
{\scaleobj{0.87}{\Omega(c^2-s^2) \frac{e^{2\Omega \tau i} }{4(1-s^2)s} \varphi(\tau)
 +s\frac{e^{2\Omega \tau i}  }{2i} \varphi'(\tau)}}
 \\[0.2cm]
 \!\!\!\!\!&=&\!\!\!\!
 {\scaleobj{0.83}{
2\frac{\Omega(c^2+s^2)}{s} id_2
 \!-\! \frac{s(2-c^2-s^2)}{1-s^2} \dot{d}_2
 \!+\!
 \Omega^2(4(1\!-\!s^2)\!-\!\frac{s^2-c^2}{(1-s^2)s^2}  ) 
 c_2 
\!-\!4s^2 \Omega i \dot{c}_2
\!+\!(s^2\!+\!1)\ddot{c}_2 }}
\\[0.39cm]
\!\!\!\!\!&&\!\!\!\!
{\scaleobj{0.87}{\Omega (2-c^2-s^2) \frac{e^{2\Omega \tau i} }{4(1-s^2)}\varphi'(\tau)
 -\Omega^2 (s^2-c^2) \frac{ e^{2\Omega \tau i} }{2i(1-s^2)s^2}\varphi(\tau)}}
 \\[0.2cm]
 \!\!\!\!\!&=&\!\!\!\!
 {\scaleobj{0.87}{4\frac{c^2 (1-s^2) \Omega^2}{s^2}  d_2
 \!-\!4c^2\Omega i \dot{d}_2 -2\frac{\Omega^3(c^2+s^2)}{s^3} i c_2
 \!+\!\frac{1-c^2 s^2}{1-s^2}\ddot{d}_2
 \!+\!\frac{\Omega^2 (2-c^2-s^2)}{(1-s^2)s} \dot{c}_2}}
 \end{array}
 \right.
 $$
 and
 $$
\left\{\!\!
\begin{array}{rcl}
 \!\!\!\!\!&&\!\!\!\! 
{\scaleobj{0.87}{\Omega(c^2-s^2) \frac{ e^{-2\Omega \tau i} }{4(1-s^2)s} \varphi(\tau)
 -s\frac{ e^{-2\Omega \tau i}  }{2i} \varphi'(\tau)}}
 \\[0.2cm]
 \!\!\!\!\!&=&\!\!\!\!
 {\scaleobj{0.83}{
-2\frac{\Omega(c^2+s^2)}{s} id_{-2}
 \!-\! \frac{s(2-c^2-s^2)}{1-s^2} \dot{d}_{-2}
 \!+\!
 \Omega^2(4(1\!-\!s^2)\!-\!\frac{s^2-c^2}{(1-s^2)s^2}  ) 
 c_{-2} 
\!+\!4s^2 \Omega i \dot{c}_{-2}
\!+\!(s^2\!+\!1)\ddot{c}_{-2} }}
\\[0.39cm]
\!\!\!\!\!&&\!\!\!\!
{\scaleobj{0.87}{\Omega (2-c^2-s^2) \frac{e^{-2\Omega \tau i} }{4(1-s^2)}\varphi'(\tau)
 +\Omega^2 (s^2-c^2) \frac{ e^{-2\Omega \tau i} }{2i(1-s^2)s^2}\varphi(\tau)}}
 \\[0.2cm]
 \!\!\!\!\!&=&\!\!\!\!
 {\scaleobj{0.87}{4\frac{c^2 (1-s^2) \Omega^2}{s^2}  d_{-2}
 \!+\!4c^2\Omega i \dot{d}_{-2} +2\frac{\Omega^3(c^2+s^2)}{s^3} i c_{-2}
 \!+\!\frac{1-c^2 s^2}{1-s^2}\ddot{d}_{-2}
 \!+\!\frac{\Omega^2 (2-c^2-s^2)}{(1-s^2)s} \dot{c}_{-2} \, . }}
 \end{array}
 \right.
 $$
Again, only the modes $m=2$ and $m=-2$ are excited by the gravitational wave. If we assume that the motion of the ring occurs purely in response to the gravitational wave perturbation, then we can set $c_m(\tau)=d_m(\tau)=0$ for $|m|\neq 2$. To solve the system of ordinary differential equations for the $m=2$ mode, we again take a Fourier transform in time. Writing
$$
\begin{array}{l}
c_2(\tau) = 
{\scaleobj{0.9}{\qfrac{1}{2\pi} \displaystyle \int_{-\infty}^{\infty}}} \widehat{c_2}(\omega) e^{i\omega \tau} d\omega \, ,
\\[0.33cm]
d_2(\tau) = {\scaleobj{0.9}{\qfrac{1}{2\pi} \displaystyle \int_{-\infty}^{\infty}}} \widehat{d_2}(\omega) e^{i\omega \tau} d\omega \, ,
\\[0.33cm]
\varphi(\tau) = {\scaleobj{0.9}{\qfrac{1}{2\pi} \displaystyle \int_{-\infty}^{\infty}}} \widehat{\varphi}(\omega) e^{i\omega \tau} d\omega \, ,
\end{array}
$$
and using again well-known result
 $
\mathcal{F}(e^{ia\tau} f(\tau)) = \widehat{f}(\omega-a) \, ,
$ 
we obtain
$$
\left\{\!\!
\begin{array}{rcl}
 \!\!\!\!\!&&\!\!\!\! 
{\scaleobj{0.84}{\tfrac{s [2 (1-s^2) (\omega-2\Omega)-\Omega ]+c^2 (\frac{\Omega}{s}) }{4 (1-s^2)}\widehat{\varphi} (\omega-2 \Omega)}}
 \\[0.2cm]
 \!\!\!\!\!&=&\!\!\!\!
 {\scaleobj{0.78}{\tfrac{(1-s^4) \omega^2-4(1-s^2) s^2 \omega  \Omega-(\frac{\Omega}{s})^2 \left(c^2+s^2 [3-4 (2-s^2) s^2]\right)}{1-s^2}\widehat{c}_{2}(\omega)\!+\!i\tfrac{s \omega  (2-c^2-s^2)-2 (1-s^2) (\frac{\Omega}{s})  (c^2+s^2)}{1-s^2}\widehat{d}_2(\omega)}} \, ,\\[0.39cm]
\!\!\!\!\!&&\!\!\!\!
{\scaleobj{0.87}{i\tfrac{(\tfrac{\Omega}{s}) [2 (\tfrac{\Omega}{s})  (s^2-c^2)+s (\omega-2\Omega)  (2-c^2-s^2)]}{4 (1-s^2)}\widehat{\varphi} (\omega-2 \Omega)}}
 \\[0.2cm]
 \!\!\!\!\!&=&\!\!\!\!
 {\scaleobj{0.78}{-i\tfrac{(\tfrac{\Omega}{s})^2 [s \omega  (2-c^2-s^2)-2 (1-s^2) (\tfrac{\Omega}{s})  (c^2+s^2)]}{1-s^2}\widehat{c}_{2}(\omega)\!+\!\tfrac{\omega^2 (1-c^2 s^2)-4 c^2 (1-s^2) [\omega \Omega+(1-s^2)(\tfrac{\Omega}{s})^2]}{1-s^2}\widehat{d}_2(\omega)}} \, .
 \end{array}
 \right.
 $$
The determinant of the system's matrix, 
$$
A(\omega)\! =\! \left[
\begin{matrix}
{\scaleobj{0.78}{\tfrac{(1-s^4) \omega^2-4(1-s^2) s^2 \omega  \Omega-(\frac{\Omega}{s})^2 \left(c^2+s^2 [3-4 (2-s^2) s^2]\right)}{1-s^2}}}
& {\scaleobj{0.78}{i\tfrac{s \omega  (2-c^2-s^2)-2 (1-s^2) (\tfrac{\Omega}{s})  (c^2+s^2)}{1-s^2}}} \\
{\scaleobj{0.78}{-i\tfrac{(\frac{\Omega}{s})^2 [s \omega  (2-c^2-s^2)-2 (1-s^2) (\tfrac{\Omega}{s})  (c^2+s^2)]}{1-s^2}}}
&
{\scaleobj{0.78}{\tfrac{\omega^2 (1-c^2 s^2)-4 c^2 (1-s^2) [\omega \Omega+(1-s^2)(\tfrac{\Omega}{s})^2]}{1-s^2}}}
\end{matrix}
\right] \, ,
$$
is given by\footnote{The root $\omega= 2 \Omega$ might be expected from the results of \cite{NQV}.}
$$
\det A(\omega) = {\scaleobj{0.9}{\qfrac{\Omega^3}{1-s^2}}}(\omega-2\Omega)
p\big({\scaleobj{0.9}{\qfrac{\omega}{\Omega}}}\big) \, ,
$$ 
where $p$ is the cubic polynomial 
$$
\begin{array}{rcl}
p(x)\!\!\!&=&\!\!\!(s^2+1) (1-c^2 s^2)x^3+2 (1-s^2) \pmb{(}1-c^2 (3 s^2+2)\pmb{)}x^2\\
&&- \pmb{(} \;\!3+c^2 (12 s^4-20 s^2+5/s^2+1)-s^2\pmb{)} x\\
&&+2 (1-s^2) \pmb{(} 1-(c/s)^2 (4 s^4-8 s^2+1)\pmb{)} \, .
\end{array}
$$
Therefore, away from the zeroes of $\det A$, we have necessarily
$$
{\scaleobj{0.9}{\bigg[
\begin{array}{r}
\widehat{c_2}(\omega) \\[0.1cm]
\widehat{d_2}(\omega) 
\end{array}
\bigg]}}
=
{\scaleobj{0.9}{\bigg[
\begin{array}{r}
(\tfrac{s}{2\Omega}) p_1(\omega/\Omega)  \\[0.1cm]
\tfrac{i}{2}\, p_2(\omega/\Omega)
\end{array}
\bigg] \qfrac{\widehat{\varphi}(\omega-2\Omega)}{p(\omega/\Omega)} \, ,}}
$$
where where $p_1$ and $p_2$ are the quadratic polynomials 
$$
\begin{array}{rcl}
p_1(x)\!\!\!\!&=&\!\!\!\! (1-s^2)\pmb{(}1+(c/s)^2(7-4s^2) \pmb{)} -(1-c^2 s^2)x^2 
\\[0.1cm]  
&&-\tfrac{1}{2}\pmb{(} (c/s)^2-(11c^2-3)-s^2+8c^2s^2 \pmb{)}x\,,\\[0.21cm]
p_2(x)\!\!\!\!&=&\!\!\!\! 2((c/s)^2+1)(s^4-3s^2+2) +\tfrac{1}{2}(s^2+1) (c^2+ s^2-2)x^2
\\[0.1cm]
&&+ \pmb{(}(c/s)^2-(3-2c^2)+2(2-c^2)s^2-2s^4\pmb{)}x \, .
\end{array}
$$

Generically, the resonant frequencies are then given by the zeroes of $p$ (whose exact expressions are too cumbersome to show here). Notice that since $p$ is a function of $\omega/\Omega$, the resonant frequencies can be tuned by adjusting the angular velocity of the ring: for example, we already know that two of these frequencies will approach $\pm \sqrt{5}ck$ (and the other will become non-resonant) as $\Omega \to 0$.

In the special case where the ring rotates with velocity equal to the longitudinal sound speed, $s\equiv \Omega R =c$, the roots of the polynomial $p$ have the simple expressions
\begin{equation} \label{3roots}
\omega= {\scaleobj{0.9}{\qfrac{2 c^2 \Omega}{1+c^2}}}, \qquad 
\omega=-{\scaleobj{0.9}{\qfrac{2(2-c^2)\Omega}{1+c^2}}}, \qquad 
\omega= 2 \Omega.
\end{equation}
In this case, the last two roots are also roots of the polynomials $p_1$ and $p_2$ (actually, $p_1(x)\equiv p_2(x)$ for~$s=c$), meaning that only the first root is resonant. This is relevant for rotating `warm' cosmic string loops (see \S\ref{section_preliminaries} and also \cite{Carter90, Vilenkin90}), which will resonate at this frequency when excited by gravitational waves. Note that such cosmic strings can be quite relativistic, that is, $s \equiv \Omega R = c$ can be of order $1$. 

For a ring rotating at a non-relativistic speed, $s\ll1$, assumed smaller than the longitudinal sound speed, $s \leq c$, the polynomials above can be approximated by\footnote{This approximation uses the fact that either $c^2 \ll 1$ or $c^2 \ll c^2/s^2$.}
\begin{equation} \label{aproxp(x)}
\begin{array}{rcl}
p(x)\!\!\!&\approx&\!\!\! x^3+2(1-2c^2) x^2-(3+5c^2/s^2)x+2(1-c^2/s^2)\,, \\[0.21cm]
p_1(x)\!\!\!&\approx&\!\!\! -x^2-\tfrac{1}{2}(3+c^2/s^2)x+1+7c^2/s^2\,, \\[0.21cm]
p_2(x)\!\!\!&\approx&\!\!\! -(1-c^2/2)x^2-(3-c^2/s^2)x+4(1+c^2/s^2)\,.
\end{array}
\end{equation}
%
The approximate expression for the polynomial $p$ matches that found by Bollada in \cite{Bol} for the resonant frequencies of a Cosserat string loop,\footnote{See equation~(41) in \cite{Bol}, where one should set $k=2$ for the mode excited by the gravitational wave, perform the substitution $x_2=2\pi\mu x$ to obtain our variable, and make the identification $\mu = s / c$, as Bollada defines the angular velocity to be $2\pi\mu$ and takes the length of the ring as the length unit and the longitudinal speed of sound as the velocity unit.} provided that we disregard the relativistic term $c^2$ in the second coefficient and also that we set $R = R_0 \equiv 1/k$ (which amounts to assuming that the radius of the rotating ring is equal to the radius of the ring in its undeformed state). This is to be expected, since the Cosserat string is the non-relativistic limit of any relativistic elastic string in the limit of small deformations (see Appendix~A in \cite{NQV}, where the Lagrangian for the Cosserat string is obtained as the Newtonian limit of a generic relativistic Lagrangian under the assumption of small deformations). Our approach is more general than that of \cite{Bol}, since we are not restricted to small deformations, and therefore $R$ will in general be larger than $R_0$, the exact relation depending on the specific elastic law of the material composing the ring (see equations~(52) and (62) and also Theorem~3.1 in \cite{NQV}). This will be especially relevant for rings constructed out of highly deformable materials. 

In the limit~$s\ll c$, the resonant frequencies are given by the simple approximate expressions 
\begin{equation} 
	\omega\approx {\scaleobj{0.9} {\pm\sqrt{5}\,\qfrac{c}{R}-\left(2-5 c^2+O(s/c)\right)\qfrac{2\Omega}{5}}}\,, \quad 
	\omega\approx {\scaleobj{0.9} {-\left(1+O\left((s/c)^2\right)\right)\qfrac{2\, \Omega}{5}}}\,,
\end{equation}
recovering the two resonant frequencies of the non-spinning ring in the limit $\Omega \to 0$, together with an additional frequency that becomes nonresonant when $\Omega=0$. Taking into account that the longitudinal sound speeds of typical materials (or even more exotic choices, such as carbon nanotubes \cite{LMW}) are in the range 1--20 km/s, we see that the relativistic corrections due to the term proportional to $c^2$ can be disregarded in this approximation (certainly thermal effects will be a far more significant concern; see equation~(101) in~\cite{TW} for an estimate of the signal-to-noise ratio as a function of the gravitational wave amplitude and the temperature and physical characteristics of the ring).

\begin{figure}[H]
	\centering
	\includegraphics[width= 9cm]{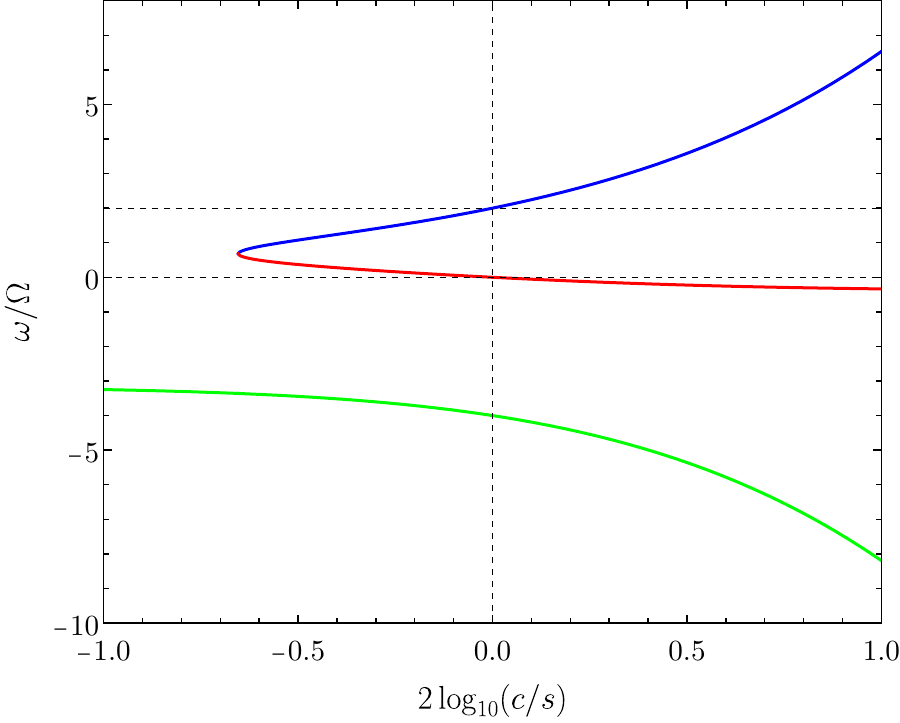}
	\phantom{AAa}
	\caption{Resonant frequencies, normalised to~$\Omega$, as a function of~$(c/s)^2$ for a nonrelativistic ring ($s,c \ll1$).}
	\label{fig:roots}
\end{figure}

Note that the approximate expressions \eqref{aproxp(x)} also hold in the regime in which~$c/s$ is of order~$O(1)$ (including~$s \ge c$) if the material is non-relativistic (i.e.,~$s,c \ll1$), in which case the corrections proportional to $c^2$ can be ignored. In Figure~\ref{fig:roots}, we show the three real roots of~$p$ as functions of~$(c/s)^2$ for nonrelativistic materials. Notice that two roots become complex for~$c/s\lesssim 0.47$, in agreement with the results of the stability analysis performed in \cite{NQV}, where it was found that elastic rings rotating with velocity~$\Omega R \equiv s \lesssim 2 c$ (in the limit~$c\ll 1$) are linearly stable. The three roots in \eqref{3roots} correspond to the red, green and blue branches, respectively. Therefore, the green and blue branches are actually not resonant for $c=s$, and the ring's response in these frequencies will be suppressed in a neighbourhood of $c=s$. 

\goodbreak 

The calculation for the mode $m=-2$ is very similar, and results in the symmetric resonant frequencies, since, as one would expect, $c_{-2}(t)=\overline{c_2(t)}$ and $d_{-2}(t)=\overline{d_2(t)}$ (in other words, the power spectrum of a real signal must be even).

An important point to have in mind is that whenever the mode $m=2$ (respectively, $m=-2$) resonates at a frequency $\omega$, it is actually responding to the frequency $\omega - 2 \Omega$ (respectively, $\omega + 2 \Omega$) of the gravitational wave. This shift of the received spectrum with respect to the emission spectrum must be taken into account when adjusting the angular velocity $\Omega$.

 \section{Conclusion}
 \label{conclusion}
 
\noindent In this work we derived the linearised relativistic elasticity equations of motion for a rod and a spinning ring encountering a gravitational wave and obtained the corresponding resonant frequencies (in the latter case extending to general elastic laws the results obtained in \cite{Bol} for spinning Cosserat strings). Both the rod and the ring were assumed to be initially lying perfectly still on a plane orthogonal to the wave's direction of propagation, so that all subsequent movement occurs in response to the perturbation introduced by the wave.

In the case of a rod with length $L$ and longitudinal speed of sound $c$, we obtained the resonant frequencies $\omega = (c/L) (\pi + 2m\pi)$ ($m\in\mathbb{Z}$) for polarisations with an axis parallel to the rod. The rod was found to be non-responsive to polarisations with an axis at an angle of $45^\circ$ with respect to the rod, so that it can be thought of as an antenna capable of tuning in to a particular polarisation.

In the case of the spinning ring, we found that only the quadrupole mode is excited by the gravitational wave. Generically, there are three resonant frequencies (counting $\pm\omega$ as the same frequency), which for nonrelativistic materials depend only on the ratio $c^2/s^2$ (where $c$ and $s$ are the ring's longitudinal and transverse speeds of sound, respectively). These three resonant frequencies become the single resonant frequency $\omega = \sqrt{5}c/R$ when the ring is not spinning (where $R$ is the ring's radius), and the single resonant frequency $\omega = 2c^2 \Omega / (1 + c^2)$ when $s=c$. All these frequencies scale with the ring's angular velocity $\Omega$, and so they can be tuned to a particular gravitational wave frequency by adjusting $\Omega$, keeping in mind that the received spectrum is shifted by $2\Omega$ with respect to the gravitational wave spectrum.

We note that, typically, an elastic ring will be resonantly excited by gravitational waves of wavelength larger than the ring radius; as an example, a nonrotating steel ring of radius $R \sim 1$~m, whose longitudinal speed of sound is $c \sim 6$~km/s, will respond to gravitational waves of frequency 
$\frac{\omega}{2\pi}\sim 2$ kHz,
corresponding to a wavelength $\lambda \sim 140$~km. Rotation will offset this frequency by amounts of the order of the angular velocity 
$\frac{\Omega}{2\pi}$, 
which is constrained by $R\Omega <\sqrt{\sigma_{\rm steel}/\rho_{\rm steel}}$, where $\sigma_{\rm steel}$ is the tensile strength of steel and $\rho_{\rm steel}$ is its density,\footnote{At this maximal angular velocity, the difference between $R$ and $R_0$ is about $1\%$.} whence $\frac{\Omega}{2\pi} \lesssim 90$~Hz. Importantly, a new resonance frequency of the order of 
$\frac{\Omega}{2\pi}$ 
will appear, allowing for the detection of 
lower frequencies. 


\section*{Declarations}

\noindent {\bf Ethics approval and consent to participate} 
Not applicable. 

\smallskip

\noindent {\bf Consent for publication} 
Not applicable. 

\smallskip

\noindent {\bf Availability of data and materials}
All data generated or analysed during this study are included in this article.

\smallskip

\noindent {\bf Competing interests} 
The authors declare no potential conflict of interest.

\smallskip

\noindent {\bf Funding} J.N. was partially supported by Funda\c{c}\~ao para a Ci\^encia e a Tecnologia (Portugal) through CAMGSD, IST-ID (projects UIDB/04459/2020 and UIDP/04459/2020), and also by the H2020-MSCA-2022-SE project EinsteinWaves, GA no. 101131233. R.V. was supported by grant no. FJC2021-046551-I, funded by MCIN/AEI/10.13039/501100011033, and by the European Union NextGenerationEU/PRTR. 
 R.V. also acknowledges the support from the Departament de Recerca i Universitats from Generalitat de Catalunya to the Grup de Recerca `Grup de F\'isica Te\`orica UAB/IFAE' (Codi: 2021 SGR 00649).

\smallskip

\noindent {\bf Authors' contributions} All authors contributed equally.

\smallskip

\noindent {\bf Acknowledgements} Not applicable.

\end{document}